\documentclass{jfm}
\usepackage{tabu}
\usepackage{graphicx,microtype,amsmath,epstopdf, epsfig,tikz,color,xcolor,ifthen}
\usetikzlibrary{math,arrows.meta}
\usepackage{xintexpr}
\usepackage[hidelinks]{hyperref}

\graphicspath{{fig_gen/fig_1/}{fig_gen/fig_2/}{fig_gen/fig_3}{fig_gen/fig_4}{fig_gen/fig_5}{fig_gen/fig_6}{fig_gen/fig_7a}{fig_gen/fig_9}{./}}

\makeatletter
\newcommand\thefontsize[1]{{#1 The current font size is: \f@size pt\par}}
\makeatother

\definecolor{purp}{rgb}{0.4,0.2,0.8}

\newcommand{\figref}[1]{Figure \ref{#1}}   
\newcommand{\tabref}[1]{Table \ref{#1}}   
\newcommand{\appref}[1]{Appendix \ref{#1}}

\newcommand{\gr}{\bnabla}                        
\newcommand{\bigO}[1]{\textit{O}\left(#1\right)} 
\renewcommand{\vec}[1]{\boldsymbol{#1}}          
\newcommand{\ten}[1]{\mathsfbi{#1}}              
\newcommand{\threeJ}[6]{\left(                   
    \begin{tabular}{ccc}
        $#1$ & $#3$ & $#5$ \\
        $#2$ & $#4$ & $#6$
    \end{tabular}
\right)}

\newcommand{\vu}{\vec{u}}                
\newcommand{\vv}{\vec{v}}                
\newcommand{\vY}{\vec{Y}}                
\newcommand{\vd}{\vec{d}}                
\newcommand{\vf}{\vec{f}}                
\newcommand{\vx}{\vec{x}}                
\newcommand{\ve}{\vec{e}}                
\newcommand{\vt}{\vec{t}}                

\newcommand{\unit}[1]{\ve_{#1}}         
\newcommand{\nhat}{\unit{n}}            
\newcommand{\zhat}{\unit{z}}            
\newcommand{\rhat}{\unit{r}}            
\newcommand{\that}{\unit{\theta}}

\newcommand{\tY}{\ten{Y}}                
\newcommand{\tT}{\ten{T}}                
\newcommand{\te}{\ten{E}}                
\newcommand{\tI}{\ten{I}}                

\newcommand{\hvu}{\widehat{\vu}}         
\newcommand{\hhvf}{\widehat{\widehat{\vf}}}         

\newcommand{\hp}{\widehat{p}}            
\newcommand{\hU}{\widehat{U}}
\newcommand{\hF}{\widehat{F}}
\newcommand{\hz}{\widehat{z}}

\newcommand{\bvu}{\overline{\vu}}             
\newcommand{\bvt}{\overline{\vt}}
\newcommand{\bvf}{\overline{\vf}}

\newcommand{\bp}{\overline{p}}                
\newcommand{\bd}{d}
\newcommand{\bz}{\overline{z}}
\newcommand{\bU}{\overline{U}}
\newcommand{\bF}{\overline{F}}

\newcommand{\htT}{\widehat{\tT}}

\newcommand{\btT}{\overline{\tT}}

\newcommand{\vPsi}{\boldsymbol{\Psi}}

\newcommand\tR{\mathsfbi{R}}

\newcommand{\vg}{\boldsymbol{g}}
\newcommand{\tg}{\mathsfbi{G}}
\newcommand{\vh}{\boldsymbol{h}}
\newcommand{\teh}{\mathsfbi{H}}
\newcommand\hu{\widehat{u}}

\newcommand\bu{\overline{u}}

\newcommand{\btR}{\overline{\tR}}

\DeclareMathOperator{\real}{re}
\usepackage{layouts}

\shorttitle{Reciprocal swimming at intermediate Reynolds number}
\shortauthor{N. J. Derr, T. Dombrowski, C. H. Rycroft and D. Klotsa}

\title{Reciprocal swimming at intermediate Reynolds number}

\author{
Nicholas J. Derr\aff{1}\corresp{\email{derr@mit.edu}},
Thomas Dombrowski\aff{2}, 
Chris H. Rycroft\aff{1}\aff{3} \and
Daphne Klotsa\aff{2}
}

\affiliation{
\aff{1}John A. Paulson School of Engineering and Applied Sciences, Harvard University, Cambridge, MA 02138, USA
\aff{2}Department of Applied Physical Sciences, University of
North Carolina at Chapel Hill, Chapel Hill, NC 27599, USA
\aff{3}Computational Research Division, Lawrence Berkeley Laboratory, Berkeley, CA, 94720
}

\begin{document}

\maketitle

\begin{abstract}
In Stokes flow, Purcell's scallop theorem forbids objects with time-reversible (reciprocal) swimming strokes from moving. In the presence of inertia, this restriction is eased and reciprocally deforming bodies can swim. A number of recent works have investigated dimer models that swim reciprocally at intermediate Reynolds numbers $\Rey \approx$ 1–-1000. These show interesting results (e.g. switches of the swim direction as a function of inertia) but the results vary and seem to be case-specific. Here, we introduce a general model and investigate the behaviour of an asymmetric spherical dimer of oscillating length for small-amplitude motion at intermediate $\Rey$. In our analysis we make the important distinction between particle and fluid inertia, both of which need to be considered separately. We asymptotically expand the Navier-Stokes equations in the small amplitude limit to obtain a system of linear PDEs. Using a combination of numerical (Finite Element) and analytical (reciprocal theorem, method of reflections) methods we solve the system to obtain the dimer's swim speed and show that there are two mechanisms that give rise to motion: boundary conditions (an effective slip velocity) and Reynolds stresses. Each mechanism is driven by two classes of sphere--sphere interactions, between one sphere's motion and 1) the oscillating background flow induced by the other's motion, and 2) a geometric asymmetry induced by the other's presence. We can thus unify and explain behaviours observed in other works. Our results show how sensitive, counter-intuitive and rich motility is in the parameter space of finite inertia of particles and fluid.

\end{abstract}

  Authors should not enter keywords on the manuscript, as these must be chosen by the author during the online submission process and will then be added during the typesetting process (see http://journals.cambridge.org/data/\linebreak[3]relatedlink/jfm-\linebreak[3]keywords.eps for the full list)

\section{Introduction}\label{sec:Intro}

The importance of how objects swim in fluids is evident in many contexts including biology, robotics, medicine and industrial applications.
From the intricate mechanisms behind flagellar swimming of bacteria that need to break time-reversibility or the design of self-propelled nanoparticles that can deliver drug cargo to cells inside the human body, to autonomous underwater vehicles that can reach the depths of the ocean for scientific expeditions or military purposes, swimming occurs across a wide range of length scales \citep{childress-book,nacht2001,vogel2008,gazzola2014}.

At small length scales, represented by small values of the Reynolds number $\Rey$, inertia is unimportant and viscous effects dominate. 
Non-inertial swimmers must use other methods to create forward motion, chiefly viscous drag asymmetry.
Due to the time-reversibility of the Stokes equations, locomotion on the microscopic scale is subject to the scallop theorem: for a swimming gait to give rise to motion, it must not consist of a time-reversible series of body deformations.
Such gaits are ``non-reciprocal.''
The scallop theorem can be restated as forbidding reciprocal swimming in the absence of inertia \citep{taylor1951,purcell1977}.
Such swimmers have received a great deal of attention.
Theoretical models include the squirmer model which was developed to represent the swimming of ciliates~\citep{Pedley16}, and slender-body theory representing swimmers as lower-dimensional thin filaments~\citep{lighthill1960}.
Other theoretical models for non-reciprocal swimmers in Stokes flow include the three-sphere swimmer~\citep{Najafi2004}, Purcell's three-linked swimmer~\citep{becker2003}, and three-body swimmers of various shapes~\citep{bet2017}.
These studies have led to classifications of Stokesian swimmers into ``pushers'' (e.g.\@ bacteria) and ``pullers'' (e.g.\@ algae) that effectively summarise the similarities and differences between swimmers across different sizes, shapes, and gaits~\citep{Lauga2009}.

In contrast, at large length scales, viscous effects can be neglected.
Inertial swimmers leverage Newton's third law to propel themselves forward by creating a backwards-directed fluid jet.
The driving swimming gaits can be reciprocal, as in the case of an oscillating rigid fin, or nonreciprocal, as in the case of a motorboat propeller.
Swimming in this regime has been the subject of much study, yielding detailed understanding of how these swimming methods scale in speed and efficiency with the properties of the swimmer and its surrounding fluid \citep{childress-book,wu2011,Hemelrijk2013,gazzola2014,Becker15,mohsen2015,gazzola2016,maertens2017}.

Between these two regimes, where $\Rey \approx 1$--$1000$, viscous forces and inertial effects are of comparable magnitude, and the equations describing swimming cannot be simplified by neglecting one or the other  \citep{vogel2008,klotsa2019}.
Investigation of swimming in this ``intermediate-$\Rey$ regime'' has typically concentrated on particular species \citep{Bartol2009,Herschlag2011,Fuiman1988ontogeny,McHenry2003}.
Examination of model swimmers can shed light on general properties of mesoscale swimming that may be used to more efficiently design and fabricate artificial swimmers \citep{park2016phototactic,feldman2021}.
Reciprocal swimmers are of particular interest, because any emergent locomotion can be strictly attributed to inertial effects.
In recent years, an asymmetric dimer has been proposed as a convenient model system for such research, because the geometry is simple and facilitates experimental, computational and analytical studies.
An important property of the dimer design is the origin of relative sphere motion.
The experimental results of \citet{Klotsa2015} show motion of an asymmetric dimer connected by a spring in a vibrated tank, similar to investigations of the motion of asymmetric bodies in oscillating flows \citep{Rednikov2004,wright2008,hector2013,Nadal2014,Collis2017,Lippera2019}.

\textcolor{black}{The dimer swimmer can also be cast as an active agent that produces its own propulsion, as would be in nature, such that the sphere separation distance oscillates not in response to external stimuli but due to internal actuation as in \figref{fig:sphere_diagram}(a).}
This system, probed extensively in work by \citet{dombrowski2019} and \citet{dombrowski2020}, shows a remarkably rich variety of behaviour for a system with a single internal degree of freedom.
At small $\Rey$, the dimer swims in the direction of the small sphere.
However, the swim speed varies non-monotonically with $\Rey$, eventually changing direction at a critical value so that the dimer swims in the direction of the large sphere.
Similar transitions were observed by \citet{Collis2017} examining a rigid dimer in an oscillating flow, where the shape and mass asymmetries could be independently tuned to give rise to two distinct transitions, and by \citet{nguyen2021}, modelling flow through avian respiratory systems.

Building on investigations of the breakdown of the scallop theorem in the presence of inertia \citet{lauga2007,lauga2011} and \citet{gonzalez-rodriguez2009} investigated the behaviour of an asymmetric dimer with large sphere densities relative to the surrounding fluid.
The oscillating spheres' motions are exactly out of phase within the non-inertial reference frame of the dimer centre of mass, but this is not true in the inertial lab frame where a phase lag is introduced.
Within the parameter space, this phase lag represents a second degree of freedom in addition to the sphere--sphere distance.
While the fluid remains Stokesian, the two degrees of freedom allow for non-reciprocal gaits and therefore net motion, effectively side-stepping the scallop theorem.

This line of research was experimentally and computationally realised in the recent work of \citet{hubert2021}, who applied a general model for bead-based swimmers \citep{ziegler2019} describing the motion of each sphere with mobility matrix coefficients.
The inertial contribution of the dense spheres is reflected in the matrix entries of the acceleration term (``mass matrix'') in a vector differential equation.
In the part of the parameter space investigated, the authors observed swimming in the direction of the small sphere for low $\Rey$.
The analysis is tractable, in part, because of the assumption of inertia-free fluid.
Fluid analysis of the (linear) Stokes equations is greatly simplified as compared to that of the (non-linear) Navier--Stokes equations.

Earlier, \citet{felderhof2016} used a similar method to describe swimming of an asymmetric dimer in an inertial fluid, capturing added mass effects with contributions to the mass matrix from the sphere and fluid densities.
However, this analysis did not capture or represent the time-averaged flow driven by Reynolds stress effects in the bulk, commonly referred to as ``steady streaming.''
Thus, the system investigated was equivalent to that of \citet{hubert2021}, with an additional contribution to particle inertial from the added mass effect.
As in \citet{hubert2021}, the analysis only showed motion in the direction of the small sphere.

\citet{Riley1966} showed that a sphere oscillating in a surrounding fluid gives rise to a time-averaged flow where, within a viscous boundary layer, fluid is drawn in towards the sphere at the poles along the axis of oscillation and ejected radially along the equator.
Outside of the layer, transport takes place in the opposite direction driven by a Reynolds stress.
Riley presented an analytical form for the flow in the limit of a large and small boundary layer (compared to the sphere radius), corresponding to the limit of large and small $\Rey$ describing the leading-order oscillation.
Numerical research has examined the non-linear streaming flow away from those limits \citep{Alassar1997,Chang1994,Chang1995,swift2009} and experimental results have recorded such flows using particle velocimetry \citep{Tatsuno1973LowRe,tatsuno1981Unharmonic,Kotas2007,Otto2008,coenen2016}.
\citet{dombrowski2019} show this reversal of flow is observed around the individual spheres in the oscillating dimer and suggest this steady streaming plays a role in the swimming direction transition.

\textcolor{black}{This system was generalised to arbitrary reciprocal surface deformations by \citet{felderhof1994inertial}.}
The authors asymptotically expanded the Navier--Stokes equations to obtain a linear equation describing a leading-order oscillatory gait-driven flow.
Knowledge of this leading-order flow suffices to calculate the swim speed arising from the lower-order steady streaming flow through the reciprocal theorem.
They showed that within such an expansion, the swim speed can be decomposed into contributions stemming from an effective steady slip velocity and a Reynolds stress in the bulk.
Finally, they formulated an eigenvalue problem relating different gaits to swimming efficiencies, describing the flow using vector spherical harmonics.
In more recent analyses, \citep{felderhof2017,felderhof2019effect}, the authors adjusted the set of vector harmonic basis functions to remain non-singular in the limit of small inertia, allowing them to probe over the full range of fluid inertia.
They observed the effects of the Reynolds stress vanish at small inertia, and that the motion arising from the Reynolds stress and steady slip nearly balance at large values of inertia.
They also observed for some gaits a switch in the swimming direction like that observed in the dimer geometry by \citet{dombrowski2019, dombrowski2020}.
Later, \citep{felderhof2021}, they updated their approach, recognising that motion of the sphere's centre of mass must be accounted for in the system force balance. 

In this work, our goal is to elucidate general principles behind reciprocal mesoscale swimming, particularly with regard to the roles of fluid and particle inertia, and to provide a unifying physical explanation \textcolor{black}{in terms of size and mass asymmetries} for the switch-like changes of direction observed in the works of \citet{Collis2017}, \citet{felderhof2017}, \citet{dombrowski2019}, and \citet{dombrowski2020}.  
We proceed as follows.
In section 2, we introduce the model swimmer, variables, equations, and parameters.
Applying an asymptotic expansion, we derive two coupled linear PDEs describing the leading-order oscillatory and steady flow.
We also use the reciprocal theorem to decompose the swim speed into contributions from an effective slip velocity and Reynolds stress.
In section 3, we describe a numerical solution using the finite element method.
The swim speed as a function of inertia is shown for representative example systems, including the dense Stokes swimmer of \citet{felderhof2016} and \citet{hubert2021}.
Finally, in section 4, we derive asymptotic scalings for the swim speed in the limit of small and large degrees of inertia, linking them to the mechanisms leading to changes in the swim direction.
We find at small $\Rey$, motion towards the smaller sphere emerges due each sphere's interaction with the oscillating background flow caused by the other.
In contrast, at large $\Rey$ this background flow becomes subdominant.
In this regime, the dominant interaction between the spheres is geometric, as for each sphere the presence of the other breaks spatial symmetry.
\textcolor{black}{The resulting motion is directed towards the more massive sphere.}

\section{Model system}\label{sec:Model}

\subsection{System parameters}

\begin{figure}
    \centering
	\includegraphics[width=0.98\textwidth]{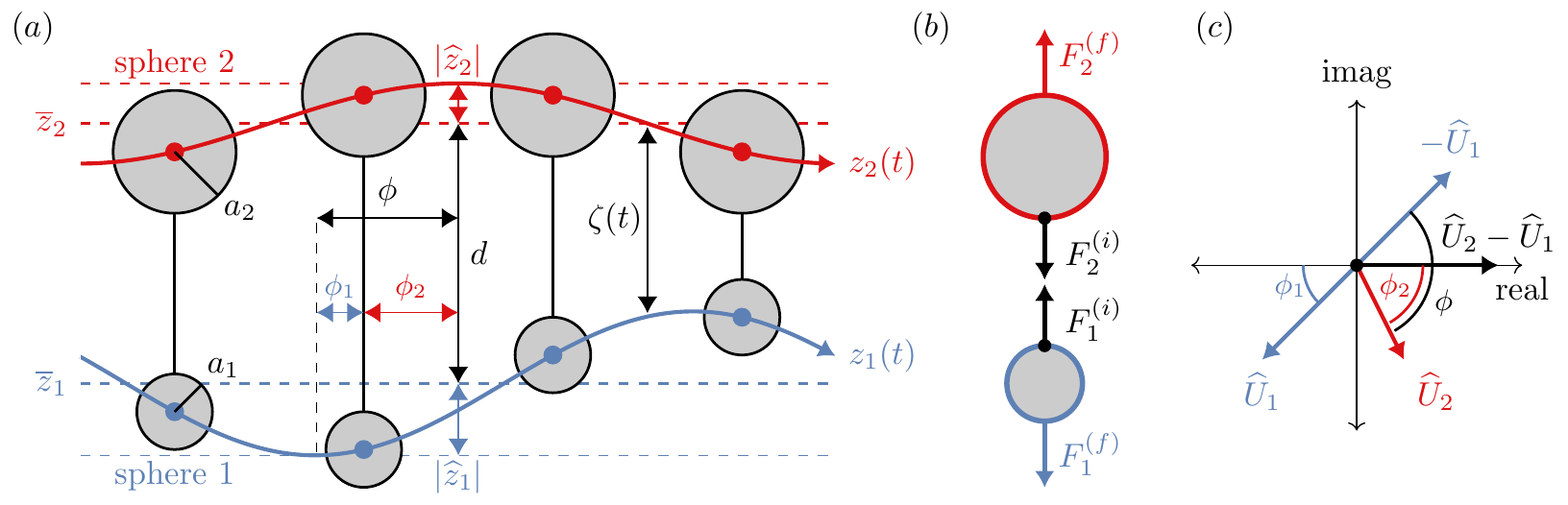}
	\caption{
    	Kinematic and dynamic model schematics.
        (a) The model system consists of a pair of spheres, labelled $j=1$ or 2, with oscillating centre positions $\smash{z_j(t) = \overline{z}_j + \hz_j e^{i\omega t}}$. 
        The distance between the spheres is $\smash{\zeta(t) = d + \Delta e^{i\omega t}}$, with $\smash{d = \bz_2 - \bz_1}$ and $\smash{\Delta=\hz_2-\hz_1}$.
        The spheres are not generally in-phase, such that sphere 1 leads the other by a phase difference $\phi$.
        (b) Each sphere is subject to two applied forces.
        The central connector exerts a pair of equal and opposite, time-dependent forces $\smash{F_2^{(i)} = -F_1^{(i)}}$, defined to produce the kinematics in (a).
        The resulting motion induces fluid-mediated forces $\smash{F_j^{(f)}}$ applied via surface tractions over the sphere surfaces.
        These include Stokes drag, the Basset force, and the added mass effect.
        (c) The spheres' velocity Fourier magnitudes $\smash{\hU_j}$ are plotted in the complex plane, illustrating the relationship between the phase difference $\phi$ and the individual phase lags $\phi_j$ between the spheres' oscillation and that of the separation rate-of-change, which has complex Fourier amplitude $\smash{i \omega \Delta = \hU_2 - \hU_1 \propto 1}$.
    }
    \label{fig:sphere_diagram}
\end{figure}

Our model system consists of two spheres submerged in a fluid of density $\rho$ and viscosity $\mu$.
We denote the fluid domain $\Omega$ and the domain of sphere $k$ as $\Omega_k$ for $k = 1$, 2.
We let the radius and density of sphere $k$ be $a_k$ and $\rho_k$, respectively.
As shown in \figref{fig:sphere_diagram}, each of the spheres is oscillating along the line connecting their centres as if connected by a massless rod of variable length applying equal and opposite forces.
We refer to the internal force, denoted with superscript $(i)$, applied to sphere 2 as $F_2^{(i)}$, and that applied to sphere 1 as $F_1^{(i)} = -F_2^{(i)}$.
While the flow induced by this motion appears time-reversible at leading order, over many cycles of oscillation small but finite inertial effects give rise to steady time-averaged drift of the two-sphere system at velocity $U$.
We seek to describe the flow velocity and pressure $\vu = \vu(\vx,t)$ and $p=p(\vx,t)$ as functions of position $\vx$ and time $t$ to precisely describe the hydrodynamic mechanisms giving rise to this steady motion and determine the form of $U$ in terms of the system parameters.

We align the $z$-axis so that it passes through both sphere centres, letting $z_k = z_k(t)$ be the instantaneous position of sphere $k$, and specifying $z_2 > z_1$ and $a_1 < a_2$ so the positive $z$-direction points toward the larger sphere.
We adopt the reference frame of the swimmer, so that the time-averaged sphere positions $\smash{{\bz}_k}$ are constants and the time-dependent velocity of each sphere $U_k = \dot{z_k}$ is periodic and zero-averaged.
In particular, we let
\begin{equation}
    U_k(t) = i\omega\hz_k e^{i \omega t}, \label{eq:U_k}     
\end{equation}
where $\omega$ is the frequency of oscillation and $\hz_k$ is the complex oscillation amplitude of sphere $k$.
Here and throughout, we implicitly take complex expressions to be equal to their real parts.
Generally, for an arbitrary time-dependent field $\psi(t)$ we will denote the zeroth and first Fourier coefficients with a bar $(\overline{\psi})$ and hat $(\widehat{\psi})$, respectively.

The positive separation distance $\smash{\zeta = z_2 - z_1}$ is thus an oscillatory function
\begin{equation}
    \zeta(t) = d + \Delta e^{i\omega t}, \qquad d = \bz_2-\bz_1, \qquad \Delta = \hz_2-\hz_1, \label{eq:sep_oscillation}
\end{equation}
where we have introduced the time-averaged separation distance $d = \overline{\zeta}$ and oscillation amplitude $\Delta = \widehat{\zeta}$ for ease of notation.
Symmetry arguments require that any steady flow must depend only on the magnitude of oscillation $\smash{|\Delta|}$, not its phase.

Without loss of generality, we assume the separation distance rate-of-change  $\smash{\frac{d\Delta}{dt}}$ has a Fourier amplitude $\smash{i\omega\Delta \propto 1}$, so that the complex arguments of the velocity Fourier amplitudes \eqref{eq:U_k} correspond to phase lags with respect to the sphere separation rate of change.
{\color{black} 
These amplitudes are given by
\begin{equation}
\hU_1 = i \omega \hz_1 = |\omega \hz_1| e^{i(\omega t + \pi + \phi_1)}, \qquad \hU_2 = i \omega \hz_2 = |\omega \hz_2| e^{i(\omega t - \phi_2)}, \label{eq:hat_U}
\end{equation}
and we define the phase difference
\begin{equation}
    \phi = \phi_1 + \phi_2 \label{eq:phi}
\end{equation}
between the spheres' extremal positions. 
The sign of $\phi$ identifies which sphere leads the other: when $\phi > 0$, sphere 1 reaches its maximum velocity before sphere 2.
When $\phi = 0$, the spheres are exactly 180$^\circ$ out of phase, and when $\phi < 0$ sphere 2 leads sphere 1.
This relationship is illustrated in \figref{fig:sphere_diagram}(c).
Shifting the argument of $\Delta$ corresponds to shifting both trajectories in \figref{fig:sphere_diagram}(a) or rotating all of the vectors in \figref{fig:sphere_diagram}(c) through the same angle, neither of which can have any effect on the time-averages obtained by integrating over a period of oscillation.
Thus, we expect at leading order $U \sim \smash{\Delta \Delta^*}$, since this is the simplest function of $\Delta$ that can be constructed which is independent of its phase.
We also expect $U$ may depend on $\phi$, $|\hU_1|$ and $|\hU_2|$, as these values are also preserved by the aforementioned shifts.
}

\subsection{Dimensionless system}

Introducing characteristic length, velocity, density, and stress scales $\smash{L= a_2}$, $\smash{V = L\omega}$, $\rho$, and $\smash{\Sigma = \mu V/L}$, we non-dimensionalise the model system.
In the following, variables and parameters should be assumed dimensionless unless otherwise specified.
Our choice of the length scale is based on the primary contribution to the Stokes drag of the dimer, and that we will explicitly represent factors of the dimensionless radius $a_2 = 1$.

The fluid system can be reduced to two non-dimensional parameters.
Firstly, we introduce $\smash{M^2 = \rho \omega a_2^2 /\mu}$ relating the time partial derivative in the Navier--Stokes equations to viscous stresses $\Sigma$.
This is the same dimensionless number and notation used to describe a single oscillating sphere by \cite{Riley1966}, although we let $M \in \mathbb{R}$ while Riley defines it as a complex number.
The quantity also appears in Stokes' second problem as the decay rate of a boundary layer with increasing height above an oscillating surface \citep{acheson1991}, and it takes the same form as the Womersley number \textit{Wo} describing pulsing flow through pipes \citep{nguyen2021}.
In settings similar to the current work, it has also been referred to as a scale parameter $s^2$ \citep{felderhof2016} and oscillatory Reynolds number $\Rey_\omega$ \citep{lauga2007}. 

Secondly, we introduce the ratio of oscillation amplitude to system size $\smash{\varepsilon = |\Delta| / L}$, which is the inverse of the Strouhal number.
The scaling of the sphere velocities \eqref{eq:U_k} shows the dimensionless sphere velocities $U_k \sim \varepsilon$, so the Reynolds number describing flow about the oscillating spheres is $\smash{\Rey = \varepsilon M^2}$.
In the following we will assume $\smash{|\Delta| \ll a_1,a_2 \ll d}$, so that the amplitude of oscillation is small and separation distance large compared to the sphere sizes.
In this regime, the flow is naturally described by the parameters $M^2$ and $\varepsilon$, but at higher oscillation amplitudes, as in the work of \citet{dombrowski2019} and \citet{dombrowski2020}, $\Rey$ and $\varepsilon$ are a more convenient set of independent parameters.

Finally, in consideration of the Stokes case, we define the parameter $S^2 = 2\rho_2 \omega a_2^2/3\mu$, representing the relative magnitude of the spheres' inertia and viscous stresses in the fluid.
The factor of 2/3 is included for comparison between the solid and fluid inertial cases.
Using this convention, in the fluid-inertial case with $\rho_1=\rho_2=1$, the effective mass (physical mass plus added mass effect) of sphere $k$ is $2 \pi M^2 a_k^3$.
In the solid inertial case $M^2 = 0$, there is no added mass effect and the mass of sphere $k$ is $2 \pi S^2 a_k^3$.
Thus, using the respective parameters for these two cases will yield a direct comparison between the effective masses of the spheres.

\def\arraystretch{1.2}
\begin{table}
    \centering
    \begin{tabular}{c|l|c}
      \raisebox{0.5em}{Symbol} & \raisebox{0.5em}{Description} & \shortstack{Definition \\ (if derived)} \\
        $a_k$ & radius of sphere $k$ & \\
        $\rho_k$ & density of sphere $k$ & \\
        $\rho$ & fluid density & \\
        $\mu$ & fluid viscosity & \\
        $\omega$ & frequency of oscillation & \\
        $\widehat{\psi}$ & first Fourier coefficient of arbitrary field or parameter $\psi = \psi(t,\dots)$ & \\
        $\overline{\psi}$ & time-averaged value of arbitrary field or parameter $\psi = \psi(t,\dots)$ & \\
        $z_k$ & time-dependent position of sphere $k$ centre & \\
        $\zeta$ & time-dependent sphere separation distance & $z_2 - z_1$\\
        $\Delta$ & first Fourier amplitude of separation distance & $\widehat{\zeta}$ \\
        $d$ & time-averaged separation distance & $\overline{\zeta}$\\
        $U_k$ & time-dependent velocity of sphere $k$ & \\
        $F_{k}^{(n)}$ & net force on sphere $k$ & \\
        $F_{k}^{(f)}$ & fluid force on sphere $k$ & \\
        $F_{k}^{(i)}$ & interior force on sphere $k$ & $F_k^{(n)} - F_k^{(f)}$ \\
        $\Rey$ & Reynolds number & $\rho \omega a_2 |\Delta| /\mu$ \\
        $M^2$ & fluid inertial parameter & $\rho \omega a_2^2/\mu $ \\
        $S^2$ & solid inertial parameter & $2\rho_2 \omega a_2^2/3\mu$ \\
        $\varepsilon$ & dimensionless amplitude of oscillation & $|\Delta|/a_2$ \\
        $\vu$ & velocity & \\
        $p$ & pressure & \\
        $\te$ & Rate-of-strain tensor & \\
        $\tT$ & Cauchy stress tensor & 
    \end{tabular}
    \caption{List of symbols used in this paper.}
    \label{tab:my_label}
\end{table}

The flow obeys the Navier--Stokes equations
\begin{equation}
  M^2 \left(\frac{\p \vu}{\p t}+ \vu \bcdot \bnabla \vu \right) = -\bnabla p + \nabla^2 \vu, \qquad \bnabla \bcdot \vu =  0, \label{eq:Navier-Stokes}
\end{equation}
subject to the boundary conditions
\begin{equation}
  \vu\left(\vx_k \right) =  U_k \zhat, \qquad \lim_{|\vx| \to \infty} \vu(\vx) = -U \zhat, \label{eq:bcs}
\end{equation}
where $\vx_k \in \partial\Omega_k$ denotes an arbitrary position on the boundary of sphere $k$.
The right hand side of \eqref{eq:Navier-Stokes} can be written as the divergence of the Cauchy stress tensor $\tT$,
\begin{equation}
    \tT = -p \tI + 2 \mu \te, \qquad \te = \frac12\left[\bnabla \vu + \left(\bnabla \vu\right)^\intercal\right],
\end{equation}
where $\tI$ is the identity tensor and $\te$ is the rate-of-strain tensor.

As shown in \figref{fig:sphere_diagram}, the spheres are subjected to a vertical flow-mediated force $F_k^{(f)}$---including contributions from the Stokes drag, Basset force and added mass effect--- in addition to the force pair $F_k^{(i)}$ applied by the central connector, given by
\begin{equation}
F_k^{(f)} = \int_{\partial\Omega_k} \zhat \bcdot \tT \bcdot \nhat dS,
\end{equation}
where $\nhat$ is the normal vector pointing into the fluid.
The net vertical force on sphere $k$ can be calculated from the acceleration as
\begin{equation}
    F_k^{(n)} = i M^2 \frac{\rho_k}{\rho} \left(\frac{4}{3} \pi a_k^3\right) U_k = i 2\pi S^2 a_k^3 \frac{\rho_k}{\rho_2} U_k, \label{eq:F_net}
\end{equation}
and at all times each sphere satisfies
\begin{equation}
    F_k^{(n)} = F_k^{(f)} + F_k^{(i)}.
    \label{eq:fnk_i}
\end{equation}
Since $F_1^{(i)} =-F_2^{(i)}$, the pair of spheres satisfies
\begin{equation}
    F_1^{(n)} - F_1^{(f)} = F_2^{(f)} - F_2^{(n)}.
    \label{eq:fnk}
\end{equation}

\subsection{Series representation}

As mentioned above, the sphere velocities \eqref{eq:U_k} scale as $\smash{U_k \sim \varepsilon \ll 1}$, suggesting the advective term in the Navier--Stokes equations $\smash{\vu\bcdot\bnabla\vu \sim \varepsilon^2}$ will be small and may, as a product of oscillatory functions, have a steady component.
As such, the flow can be described as a double sum over powers of $\varepsilon$ and Fourier modes.
\textcolor{black}{
See \appref{app:ansatz} for a detailed derivation. The approach follows that of \citet{felderhof1994inertial}, who applied it to a single-sphere geometry.
}
Ignoring all terms with no effect on the $\bigO{\varepsilon^2}$ swim speed, we adopt the ans\"atz
\begin{equation}
    \vu = \varepsilon \hvu e^{it} + \varepsilon^2 \bvu , \qquad p = \varepsilon \hp e^{it} + \varepsilon^2 \bp, \label{eq:ansatz}
\end{equation}
so that $(\hvu, \hp)$ are complex fields describing the Fourier amplitudes of a leading-order oscillatory flow and $(\bvu,\bp)$ the steady flow field generated by inertial effects as described above.
A $\pi$-periodic flow is also induced at the same order as the steady flow, which it does not influence. 
\textcolor{black}{For this reason we omit its description as in the work of Felderhof \& Jones cited above.}
Since we have chosen an inertial reference frame, $U = \varepsilon^2 \overline{U}$.
Similarly, the periodic sphere velocities are decomposed as $U_k(t) = \varepsilon \hU_k e^{it}$, where $\hU_k$ are the non-dimensionalisation of the complex Fourier amplitudes \eqref{eq:hat_U}, implying the net force is $F_k^{(n)} = \varepsilon \hF_k^{(n)} e^{i t}$.
The convention of using hats and bars to denote first and zeroth Fourier amplitudes should now be understood to include this normalisation by $\varepsilon$ and $\varepsilon^2$, respectively, so that the amplitudes are $\bigO{1}$.

\textcolor{black}{
The form of \eqref{eq:ansatz} shows a potential inconsistency: if $\bU$ grows large at high $M^2$, there may be steady flow at order-1 or even order-0, violating the assumptions under which the ans\"atz was introduced.
However, it is shown in section 4 that the swim speed $\bU$ approaches a finite value $\bU_\infty$ as $M \to \infty$ rather than growing unboundedly.
Other, similar analyses in the one-sphere geometry \citep{felderhof1994inertial,felderhof2017} have also shown the steady flow remains bounded at high inertia.
We note that the ans\"atz analysis does break down at large enough $M^2$ or $\varepsilon$.
However, this is due to turbulent effects which are not represented in this laminar description as opposed to an inconsistent set of assumptions.
}

Substituting the expansion \eqref{eq:ansatz} into the Navier--Stokes equations \eqref{eq:Navier-Stokes} shows the Fourier amplitudes are described by
\begin{equation}
    \left(\nabla^2 - i M^2\right)\widehat{\vu} = \bnabla \widehat p, \qquad \gr \bcdot \hvu = 0.  \label{eq:Brinkman}
\end{equation}
These are a complex version of the Brinkman equations, a combination of the Stokes equations and Darcy flow where both viscous stresses and frictional drag force are non-negligible \citep{durlofsky1987}.
In this case, the ``drag term'' $-iM^2 \vu$ is imaginary, arising not from the effects of some porous structure but from the acceleration of the unsteady fluid 90$^\circ$ out of phase.
The oscillation gives rise to a boundary layer of width $1/M$ over which the resulting vorticity is diffused \citep{Riley1966}.

The steady component of the corresponding advective term can be interpreted as a Reynolds stress $\btR$ driving the second-order steady flow according to the Stokes equations,
\begin{equation}
     \nabla^2 \overline{\vu} = \bnabla\overline p - \bnabla \bcdot \btR, \qquad \btR = -\frac{M^2}{2} \ \hvu \otimes \hvu^*, \label{eq:Stokes}
\end{equation}
where the asterisk denotes a complex conjugate.
The factor of 1/2 in the Reynolds stress arises from considering the real part of a product of complex exponentials, as described in \appref{app:ansatz}.

The sphere boundaries $\p\Omega_k$ are moving.
To obtain a time-independent system, we derive flow constraints on the time-averaged boundaries $\overline{\p\Omega}_k$ \textcolor{black}{by Taylor expanding the flow fields about points on this surface and matching terms at each order in the ans\"atz.}
This process, described in \appref{app:ansatz}, yields boundary conditions on a static geometry for the Brinkman amplitudes
\begin{equation}
  \hvu(\vx)|_{\vx \in \overline{\p\Omega}_k} = \hU_k \zhat, \qquad \lim_{|\vx|\to \infty}\hvu(\vx) = \vec{0} \label{eq:Brinkman_bc}
\end{equation}
and steady flow
\begin{equation}
\bvu(\vx) |_{\vx \in \overline{\p\Omega}_k} = \bu_s \that := \frac{i\hU_k}{2}\frac{\p\hvu^*}{\p z}, \qquad \lim_{|\vx| \to \infty} \bvu(\vx) = -\overline{U} \zhat. \label{eq:Stokes_bc}
\end{equation}
We have introduced the steady tangential slip velocity $\bu_s = \bu_s(\theta)$ defined on $\overline{\p\Omega}_{k}$, where $\theta$ is the polar angle from the $z$-axis in spherical coordinates originating at the centre of sphere $k$.
\textcolor{black}{We emphasise that this slip velocity is unrelated to the Brinkman boundary layer described above, instead arising as in the ``swimming sheet'' of \citet{taylor1951} from periodic motion of the boundary.}

Writing $\hvu = \hu_r \rhat + \hu_\theta \that$ and $\bvu = \bu_r \rhat + \bu_\theta \that$ (letting $r,\theta$ and their associated unit vectors refer to the spherical coordinates at sphere $k$), we briefly show that $\p_z \hvu \propto \that$ on sphere $k$  as claimed.
Since $\p_\theta \hvu = 0$ on the sphere surface, $\p_z\hvu = \cos\theta \p_r \hvu$, implying $\bvu_r \propto \p_r \hu_r$.
The divergence-free condition requires $\p_r \hu_r = -(2 \hu_r + \cot \theta \hu_\theta + \p_\theta \hu_\theta)/r$.
Substituting in $\hu_r = \hU_k \cos\theta$ and $\hu_\theta = -\hU_k \sin \theta$ shows this quantity vanishes and $\bu_r = 0$ on the surface.
\textcolor{black}{Thus, the order-$\varepsilon^2$ steady flow field $\bvu$ obeys the physical requirement of no flux through sphere surfaces.}

The swim speed $\bU$ is unknown, as are the individual sphere oscillation amplitudes $\hU_1$ and $\hU_2$.
We introduce
\begin{equation}
    \widehat{F}_k^{(f)} := \int_{\overline{\p\Omega}_k} \zhat \bcdot \htT \bcdot \nhat \ dS, \qquad \overline{F}_k^{(f)} := \int_{\overline{\p\Omega}_k} \zhat \bcdot \btT \bcdot \nhat \ dS,
\end{equation}
where $\htT$ and $\btT$ are the Cauchy stress tensors corresponding to the oscillatory and steady flows $\left(\widehat{\vu},\widehat p\right)$ and $\left(\overline{\vu},\overline p\right)$, respectively.
The three unknowns are thus fixed by the constraints
\begin{equation}
	\hU_2 - \hU_1 = 1, \qquad \hF_{i,1} + \hF_{i,2} = 0, \qquad \bF_{f,1} + \bF_{f,2} = 0.
\label{eq:constraints}
\end{equation}
See \appref{app:ansatz} for a complete derivation of the boundary conditions and force conditions.

Due to linearity, the steady solution $\bvu = \bvu_b + \bvu_r$ can be decomposed into flows driven exclusively by the boundary condition ($\bvu_b,p_b$) and Reynolds stress ($\bvu_r,p_r$), respectively, such that
\begin{equation}
	\nabla^2 \bvu_b = \bnabla \bp_b, \qquad \bvu_b(\vx) \Big|_{\vx \in \overline{\p\Omega}_k} = u_s \that, \qquad \lim_{|\vx| \to \infty} \bvu_b = -\bU_b \zhat, \label{eq:stokes_b}
\end{equation}
\begin{equation}
	\nabla^2 \bvu_r = \bnabla \bp_r - \bnabla \bcdot \btR, \qquad \bvu_r(\vx) \Big|_{\vx \in \overline{\p\Omega_k}} = \vec{0}, \qquad \lim_{|\vx| \to \infty} \bvu_r = -\bU_r \zhat, \label{eq:stokes_r}
\end{equation}
where the mechanism-specific swim speeds $\bU_b$ and $\bU_r$ satisfy $\bU = \bU_b + \bU_r$.
In particular, by applying the reciprocal theorem as in \appref{app:reciprocal}, the two speeds can be written as
\begin{equation}
    \bU_b = \sum_{k=1}^2 \frac{i \hU_k}{2F'} \int_{\overline{\p\Omega}_k} \nhat \bcdot \tT^\prime \bcdot  \frac{\p \hvu^*}{\p z} dS, \qquad \bU_r = \frac{M^2}{2 F'} \int_\Omega \hvu \bcdot \te^\prime \bcdot \hvu^* dV, \label{eq:swim_speeds}
\end{equation}
where the primed variables correspond to the Stokes flow resulting from towing the dimer at a velocity $U^\prime$ with a force $F^\prime$.
The full set of model equations \eqref{eq:Brinkman}--\eqref{eq:constraints} may be solved numerically using the finite element method and analytically with the method of reflections.
In the next section, we use the finite element approach to examine the swim speed as a function of $M^2$ or, when considering a dense swimmer in Stokes flow as in \citep{hubert2021}, a function of $S^2$.
In Section 4, we interpret these results analytically.

\section{Numerical treatment}
\subsection{Finite element method}
The dimensionless Brinkman equations, given a parameter $\alpha$ representing an inverse screening length, are written
\begin{equation}
	\left(\nabla^2 - \alpha^2\right) \vu = \bnabla p - \vf, \qquad \bnabla \bcdot \vu = 0.
\label{eq:real_Brinkman}
\end{equation}
They describe flow subject to a body force $\vf \in L^2(\Omega)$ in a porous region where a drag force $-\alpha^2 \vu$ is of the same magnitude as viscous stresses.
If $\alpha = 0$, they are the Stokes equations.
Therefore, we develop our numerical solution procedure for arbitrary complex $\alpha$.

If $\vu$ has Dirichlet boundary conditions, it is well known \citep{iliev2011} that \eqref{eq:real_Brinkman} has a unique weak solution $\left(\vu, p\right) \in H^1(\Omega)^2 \times L_0^2(\Omega)$ such that for all test functions $(\vv,q) \in H^1(\Omega)^2 \times L_0^2(\Omega)$,
\begin{multline}
	\int_\Omega \left[ \left(\gr \vv \vec{:} \gr \vu\right) + \alpha^2 \left(\vv \bcdot \vu\right) - (\gr \bcdot \vv) p\right] dV = \int_{\Omega} \vv \bcdot \vf \ dV, \qquad  \int_{\Omega} q(\gr \bcdot \vu) dV = 0.
\label{eq:Brinkman_weak}
\end{multline}
At this point we discretise our axisymmetric domain into two $n_e \times n_e$ grids of curvilinear, quadrilateral elements using the bispherical coordinates $(\xi,\eta)$.
The surface of sphere $k$ is given by $\xi = \xi_k$, with $\xi_1 < 0 < \xi_2$, and by symmetry the system is agnostic to the substitution $\eta \to 2\pi - \eta$. 
The grids are periodic over $\eta \in [0,2\pi)$, spanning $\xi \in [\xi_1,0]$ and $\xi \in [0,\xi_2]$, respectively.
We denote the collection of elements $\mathcal{T}_{n_e}$.
On each element, we consider functions defined within the space \mbox{$Q_m = \text{span} \left\{\xi^j \eta^k : 0 \le j,k \le m\right\}$}, and globally we consider piecewise combinations of these.
In particular, we let $\mathcal{C}^1_m\left(\mathcal{T}_{n_e}\right)$ be the space of such functions that are globally continuous.
We seek approximate solutions \mbox{$\smash{(\vu^h,p^h) \in C^1_{n_p}\left(\mathcal{T}_{n_e}\right)^2 \times C^1_{n_p-1}\left(\mathcal{T}_{n_e}\right)}$}, i.e.
defined on the so-called generalised Taylor-Hood $\smash{\left(Q_{n_p}\text{- }Q_{n_p-1}\right)}$ element.
In terms of basis functions for vector and scalar fields, $\smash{\vPsi_k(\vx) \in C^1_{n_p}\left(\mathcal{T}_{n_e}\right)^2}$ and $\smash{\psi_k(\vx) \in C^1_{n_p-1}\left(\mathcal{T}_{n_e}\right)}$, we write
\begin{equation}
	\vu^h = \sum_j^{N_u} u_j^h \vPsi_j, \qquad p^h = \sum_j^{N_p} p_j^h \psi_j,
\end{equation}
where $N_u$ and $N_p$ are the number of velocity and pressure degrees of freedom, respectively.
Considering test functions $(\vv^h,q^h) \in C^1_{n_p}\left(\mathcal{T}_{n_e}\right)^2\times C^1_{n_p-1}\left(\mathcal{T}_{n_e}\right)$ and substituting into \eqref{eq:Brinkman_weak}, the coefficients $u_j^h$ and $p_j^h$ are known to satisfy the linear system
\begin{equation}
	\left(\begin{tabular}{cc} $A(\alpha^2)$ & $-B^\mathsf{T}$ \\ $B$ & 0 \end{tabular}\right) \left(\begin{tabular}{c} $u^h$ \\ $p^h$ \end{tabular}\right) = \left(\begin{tabular}{c} $F$ \\ 0\end{tabular}\right), \label{eq:FEM}
\end{equation}
where the block matrix elements are given by
\begin{equation}
	A_{ij}(\alpha^2) = \int_\Omega \left(\gr \vPsi_i \vec{:} \gr \vPsi_j + \alpha^2 \vPsi_i \bcdot \vPsi_j\right) dV, \qquad B_{ij} = \int_\Omega \psi_i \left(\gr \bcdot \vPsi_j\right)  dV,
\end{equation}
and the source term on the right-hand side is
\begin{equation}
	F_i = \int_\Omega \vPsi_i \bcdot \vf dV.
\end{equation}
In bispherical coordinates, the velocity boundary conditions can be written 
\begin{equation}
	\vu(\xi_k,\eta) = V_k \zhat + v_s(\eta)\that, \qquad \vu(0,0) = -V,
\end{equation}
for a set of sphere velocities $V_k$, a boundary slip velocity $v_s$, and the velocity $V$ of the swimmer frame relative to the lab frame.
We use $V$ and $v$ to distinguish numerical parameters from system parameters.
In all cases below, we directly solve \eqref{eq:FEM} using the distributed-memory version of the SuperLU library \citep{li05} and PETSc library for scientific computation \citep{petsc-efficient,petsc-user-ref,petsc-web-page}, working in C\texttt{++}.
The solution procedure is as follows.

\subsection{Solution procedures} 
\subsubsection{Brinkman equations}
The Brinkman equations \eqref{eq:Brinkman} correspond to the linear system \eqref{eq:FEM} with $\smash{\alpha^2 = iM^2}$, $\smash{\vf = \vec{0}}$, $\smash{V_k = \hU_k}$, $\smash{v_s = 0}$, and $V=0$.
The unknown numerical solution $(\vu^h,p^h)$ are complex fields, stemming from the complex symmetric  block matrix $A$.
Recall the sphere amplitudes $\hU_k$ are unknown and fixed by the constraints \eqref{eq:constraints}.
To find them, we set $\smash{V_k = \tilde{U}_k}$, trial values of the sphere velocity amplitudes, which we consider as a vector in $\mathbb{C}^2$.
These yield a corresponding set of force amplitudes $\smash{\tilde F_k^{(i)}\in \mathbb{C}^2}$ upon solution of \eqref{eq:FEM}. 

Since the equations are linear, we have $\smash{\tilde{F}_i^{(k)} = C_{kj} \tilde{U}_j}$ for some constant complex matrix $\smash{C \in \mathbb{C}^{2 \times 2}}$, with $\smash{C_{kj} = d\tilde{F}_i^{(k)}/d\hU_j}$.
In our approach, we calculate $\smash{\tilde{F}_i^{(k)}(\tilde{U}_j)}$ at $\smash{(\tilde{U}_1,\tilde{U}_2) = (1,0), \ (0,1), }\text{ and }\smash{(0,0)}$, and we use the results to compute $\smash{d\tilde{F}_i^{(k)}/d\tilde{U}_j}$ at $\smash{(0,0)}$ via finite difference.
The constraints \eqref{eq:constraints} require $\smash{\hF_i^{(1)} = -\hF_i^{(2)}}$ and $\smash{\hU_2 - \hU_1 = 1}$.
From the problem definition, $\smash{\tilde{F}_i^{(k)} = 0}$ at $\smash{(\tilde{U}_1,\tilde{U}_2)=0}$, so we set
\begin{equation}
	\hU_k = \frac{\hU_{k,*}}{\hU_{2,*}-\hU_{1,*}}, \qquad \hU_{k,*} = C^{-1}_{kj} \tilde{F}_{j,*}, \qquad \tilde{F}_{2,*} = -\tilde{F}_{1,*} = 1, \label{eq:unknown_U}
\end{equation}
for an intermediate set of force and velocity amplitudes $\smash{\tilde{F}_{k,*}}$ and $\smash{\hU_{k,*}}$.
We solve \eqref{eq:FEM} once more with this set of velocity amplitudes to obtain a solution that satisfies \eqref{eq:constraints}. 

\subsubsection{Stokes equations}
The Stokes equations \eqref{eq:Stokes} correspond to \eqref{eq:FEM} with $\alpha^2 = 0$, $\vf = \gr \bcdot \btR$,  $V_k = 0$, $v_s = \bu_s$, and $V = -\bU$, where the quantities $\btR$ and $\bu_s$ are computed by post-processing the solution above.
Now, the unknowns $(\vu^h,p^{h})$ are real fields, since $A$ and $F$ are real.
As before, $\bU$ is not known \textit{a priori}, and must be calculated to satisfy $\bF = 0$ by initially solving with a trial value $V = -\tilde U$.
 Let $\tilde{F}$ now denote the calculated value of $\bF$ given the velocity input $V = -\tilde U$.

By the equations' linearity, $\tilde{F} = D \tilde{U} + \tilde{F}_0$ for some constants $D = d\tilde{F} / d\tilde{U}$ and $\tilde{F}_0 = \tilde{F}(0)$, the steady force at $\tilde{U} = 0$.
We compute the two constants by solving the system for $\tilde{U} = 0$ and $\tilde{U} = 1$, calculating $\tilde{F}$ for both.
The former gives $\tilde{F}_0$ directly; we calculate $D$ via finite difference as before.
Finally, we let $\bU = -\tilde{F}_0/D$.
Solving \eqref{eq:FEM} once more gives the steady flow field $(\bvu,\bp)$.
Note that we can calculate $\bU_b$ or $\bU_r$ in isolation by altering the parameters in accordance with \eqref{eq:stokes_b} or \eqref{eq:stokes_r} as necessary.

\subsection{Results}
Using the procedure above, we solve for the steady flow $\bvu$ and associated swim speed $\bU$ as a function of inertia.
We also calculate the mechanism-specific speeds $\bU_b$ (boundary condition) and $\bU_r$ (Reynolds stress) for three characteristic examples.
In all cases we let $a_1 = 1/2$ and $\bd = 3$.
Recall $a_2 = 1$ due to the chosen length scale.

\subsubsection{Visualisation of steady flow field}
\begin{figure}
    \centering
	\includegraphics{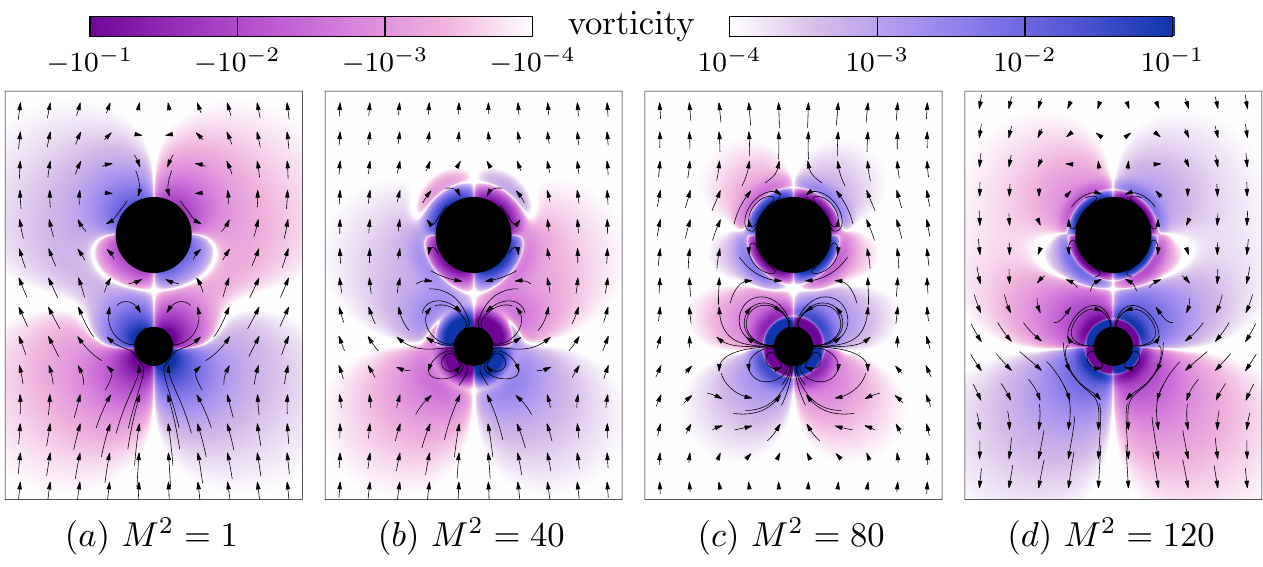}
    \caption{The steady flow field as a function of fluid inertia $M^2$, plotted via streamlines in the reference frame of the swimmer.
Vorticity is denoted with colour. 
(a) At small $M^2$ the dimer acts as a ``puller'', attracting fluid along its axis of symmetry and ejecting it radially as it swims in the direction of the small sphere.
This produces an approximately Stokesian straining flow.
(b--c) For intermediate $M^2$, there exists a boundary layer outside of which the flow is noticeably weaker.
As $M^2$ grows, the inner layer grows thinner.
(d) At large $M^2$, the dimer swims toward the large sphere and acts as a pusher.
Parameters: $(a_1,a_2,\bd,\rho_1,\rho_2) = (1/2,1,3,1,1)$.}
    \label{fig:steady_flows}
\end{figure}

First, we consider the steady flow field of a dimer with the same density as the surrounding fluid, $\rho_1=\rho_2 = 1$, as studied by \citet{dombrowski2019}.
The streamlines depict flow in the reference frame of the swimmer.
\figref{fig:steady_flows}(a) shows the system at $M^2 = 1$,  with a set of vortex rings detectable around each sphere as predicted by \citet{Riley1966} in the case of a single sphere.
Note the dimer is moving towards the small sphere in the lab frame.
In this regime, the dimer acts as a puller, attracting fluid along its axis of symmetry and ejecting it radially.
As $M^2$ increases, a boundary layer develops.
Panels (b) and (c) show its width decrease, and the development of inner and outer vorticity regions of opposing sign. 
The corresponding set of double vortex rings was also calculated by \citet{Riley1966} in the high-$M^2$ limit.
Finally, in panel (d), the swim direction has switched.
Below, we will calculate the swim speed's functional form for the $\rho_1=\rho_2=1$ case and two others.
Ultimately, we will interpret the results using asymptotic scalings of $\bU_b$ and $\bU_r$ for small and large inertia which are analogous to Riley's limits.

\subsubsection{Inertial dependence of the swim speed}
\begin{figure}
	\centering
	\includegraphics{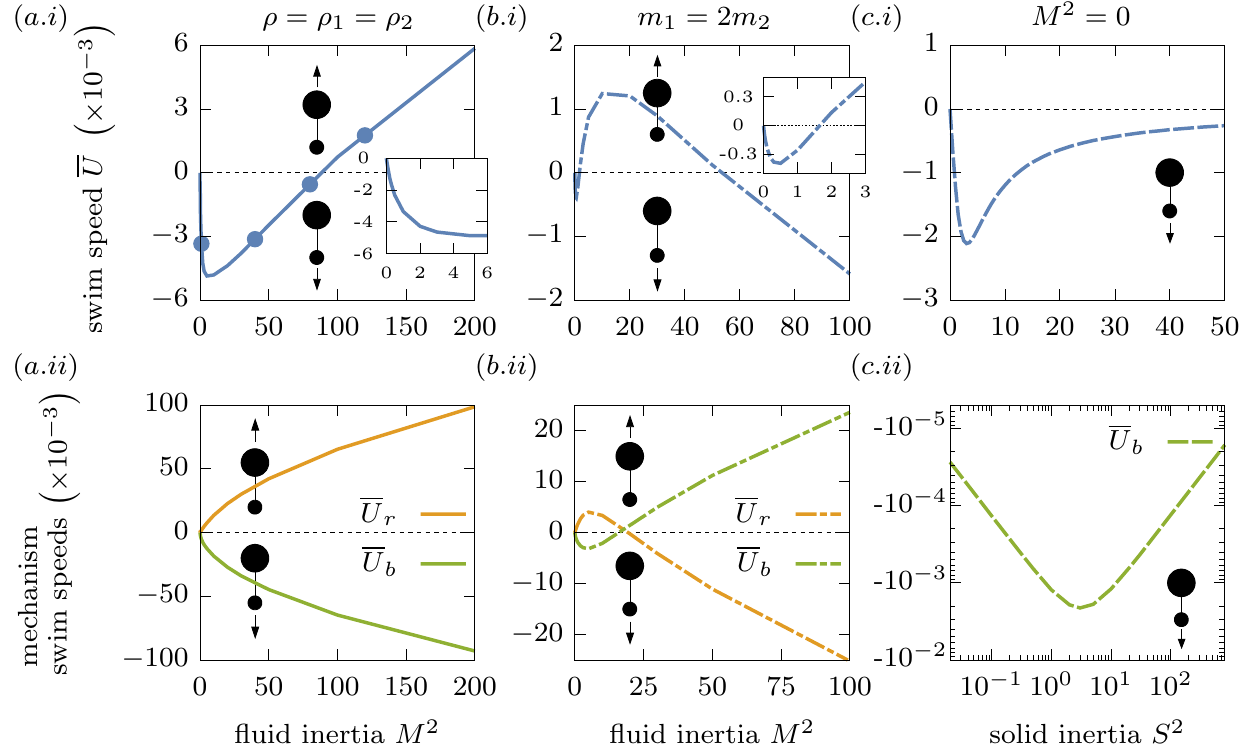}
	\caption{Survey of net dimer swim speeds (a--c.i) and contributions from each mechanism (a--c.ii) for three representative systems.
In all cases, $\{a_1,a_2,\bd\} = \{1/2,1,3\}$.
(a) For equal-density spheres and fluid $\rho_1=\rho_2 = 1$, the dimer undergoes a switch in direction when the contribution from the Reynolds stress ($\bU_r$), driving motion towards the large sphere, overcomes that of the boundary velocity ($\bU_b$), which favours motion towards the small sphere. In (a.i), the blue dots correspond to the flow fields presented in \figref{fig:steady_flows}. 
(b) If $m_1 > m_2$ the dimer may experience two direction switches as shown here, and the mechanism-specific swim speeds $\bU_b$ and $\bU_r$ are not monotonic.  (c) In Stokes flow ($M^2 = 0$) with $\rho_1 = \rho_2$, there is no change in swimming direction, and no motion at large $S^2$.}

	\label{fig:first_speeds}
\end{figure}

In \figref{fig:first_speeds}, we show the functional form of three characteristic model systems vs the degree of inertia in the system.
The first row (a--c.i) shows the overall swim speed; the second (a--c.ii) shows the mechanism-specific speeds.

In the first column (a), we consider the equal-density case $\rho_1=\rho_2=1$, where the two mechanisms drive motion in opposite directions.
\textcolor{black}{The blue dots in panel (a.i) correspond to the four steady flows shown in \figref{fig:steady_flows}.}
At low levels of fluid inertia, the swim speed is dominated by the boundary velocity and $\bU\approx \bU_b$.
At high levels of fluid inertia, $|\bU_r| \approx |\bU_b|$, and the total swim speed magnitude is much lower than that of either mechanism in isolation.
The results indicate the direction of motion changes at a critical level of inertia, as in the full Navier-Stokes simulations of \citet{dombrowski2019,dombrowski2020}.
The swim speed calculated here and shown in \figref{fig:steady_flows}(a) has the same qualitative shape as that work.
One difference to note is that there, the direction switch occurs at a critical value of $\Rey = \varepsilon M^2$ across different systems.  

\textcolor{black}{This change of direction was also observed in the single-sphere swimmer investigated by \citet{felderhof2017} and in rigid dimers in an externally oscillated flow \citep{Collis2017}.
Later, \citet{felderhof2021} showed the swim direction switch depended on the particular choice of surface deformation stroke as defined by a vector of multipole expansion coefficients.
In the current work, the number of degrees of freedom of the stroke is much smaller, depending on the relative size and mass of the two spheres.
It will be shown that low- and high-inertial limiting swim speeds can be understood as functions of the size and mass (respectively) asymmetries of the dimer.
}

In the second column (b), we consider a system where the smaller sphere is heavier than the larger one.
We introduce the effective mass of sphere $k$,
\begin{equation}
m_k = a_k^3 (\rho_k + 1/2), \label{eq:eff_mass}
\end{equation}
including the sphere's actual mass and the added mass effect caused by accelerating the surrounding fluid.
For this case, we choose $\rho_1$ and $\rho_2$ so $m_1 = 2 m_2$ with $\rho_1 a_1^3 + \rho_2 a_2^3 = \rho(a_1^3 + a_2^3)$, such that the dimer is neutrally buoyant.
While we again observe motion towards the small sphere at $M \ll 1$, the direction of swimming changes much sooner than in the previous case before switching again near the previous critical $M^2$.
In (b.ii), we see each mechanism changes direction as well.
Note that, consistent with (a), the dimer moves towards the small sphere at small $M^2$ and towards the more massive sphere at large $M^2$.
This is similar to the double switch in direction observed by \citet{Collis2017}.

In (c), the same system is shown without the effects of fluid inertia ($M^2 = 0$), so that $\bU$ is shown as a function of the solid inertia $S^2$ for $\rho_1 = \rho_2$.
This is the case investigated by \citet{hubert2021}.
\textcolor{black}{It is also mathematically equivalent to the model of \citet{felderhof2016}, which neglected the effects of Reynolds stress.
}
By definition, $\bU_r = 0$, so in the second column rather than re-plotting $\bU=\bU_b$ we instead plot the function in log--log space.
There are no changes in direction, and we observe that $\bU \to 0$ as $S^2 \to \infty$ in contrast to the other cases where $\bU$ diverges as $M^2 \to \infty$.
Below, we will see this functional form is proportional to the quantity imag$\{\hU_1 \hU_2^*\}$, indicating the swim speed is determined by interactions between the oscillating flow fields of each sphere.
\textcolor{black}{The functional form in (c.i) is identical to the swim speed presented by \citet{felderhof2016} when the effective sphere masses (including the added mass effects of accelerating the surrounding fluid) are substituted for the spheres' actual masses.}
Next, we consider how those effects enter the problem as one considers small values of the quantity $M^2 / S^2$.

\begin{figure}
    \centering

	\includegraphics[width=\textwidth]{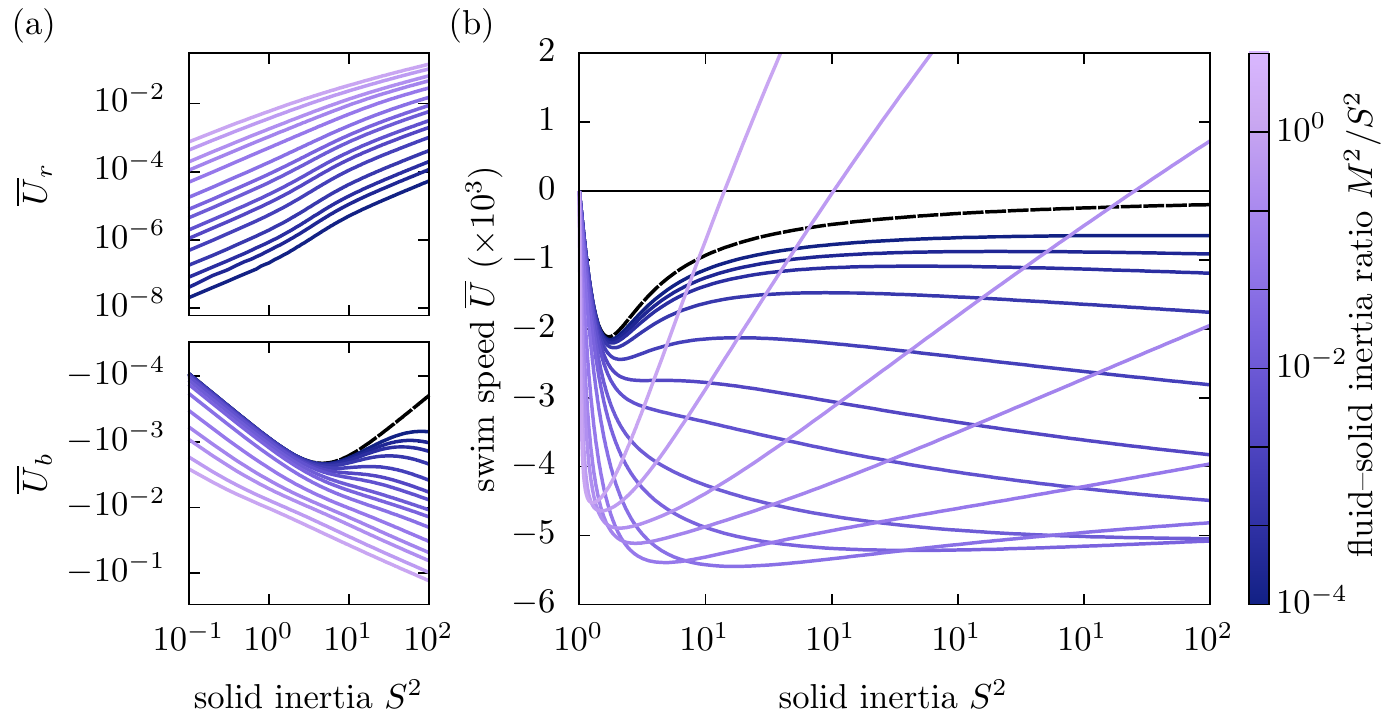}
    
	\caption{Swim speeds as a function of solid inertia for varying fluid--solid inertia ratios.
The $M^2 = 0$ case is plotted as a black dashed line, and $\rho_1 = \rho_2$ for all cases.
(a) Contributions to the swim speed from the boundary velocity $(\bU_b)$ and Reynolds stress $(\bU_r)$ are plotted as a function of the solid inertia $S^2$ for a variety of $M^2/S^2$.
Note $\bU_r = 0$ for $M^2 = 0$.
(b) The dimer swim speed is plotted for the same fluid--solid inertia ratios.
Parameters: $\{a_1,a_2,\bd\} = \{1/2,1,3\}$.}
    \label{fig:solid-to-general}
\end{figure}

In \figref{fig:solid-to-general}, we consider the functional form of $\bU$, $\bU_b$ and $\bU_r$ with $S^2$ for varying values of the ratio $M^2 / S^2$.
When this value is 0, we recover Stokes flow, plotted with a dashed black line.
Panel (a) shows the boundary velocity and Reynolds stress contributions to the swim speed behave qualitatively differently with respect to $S^2$.
While $\bU_r = 0$ for $M^2 = 0$, it is nonzero and monotonically increasing in the presence of any fluid inertia.
In contrast, $\bU_b$ is non-monotonic for $M^2 \ll S^2$.
Its magnitude increases to a peak near $S^2 \sim 1$ before beginning to decrease.
If $M^2 = 0$, it attenuates completely, but for $M^2 > 0$ the contribution ultimately begins increasing again for $S^2 \gg 1$.
Note that if $M^2 / S^2$ is large, this non-monotonicity is not detected.
\textcolor{black}{Plots of the total swim speed (b) suggest that $\bU$ does not vanish as $S^2 \to \infty$ for any $M^2 > 0$.}

\section{Analytical treatment}

Although solving the Stokes equations \eqref{eq:Stokes} is necessary to obtain the steady flow field $\bvu$, the swim speed is completely determined by the Brinkman amplitudes $\hvu$ \eqref{eq:Brinkman} as can be seen from the form of \eqref{eq:swim_speeds}.
As such, in order to interpret the results above, we turn to the axisymmetric motion of two spheres in a Brinkman medium to describe the various swim mechanisms analytically.

\subsection{Describing flow about two spheres}

\begin{figure}
	\centering
	\includegraphics[width=0.85\textwidth]{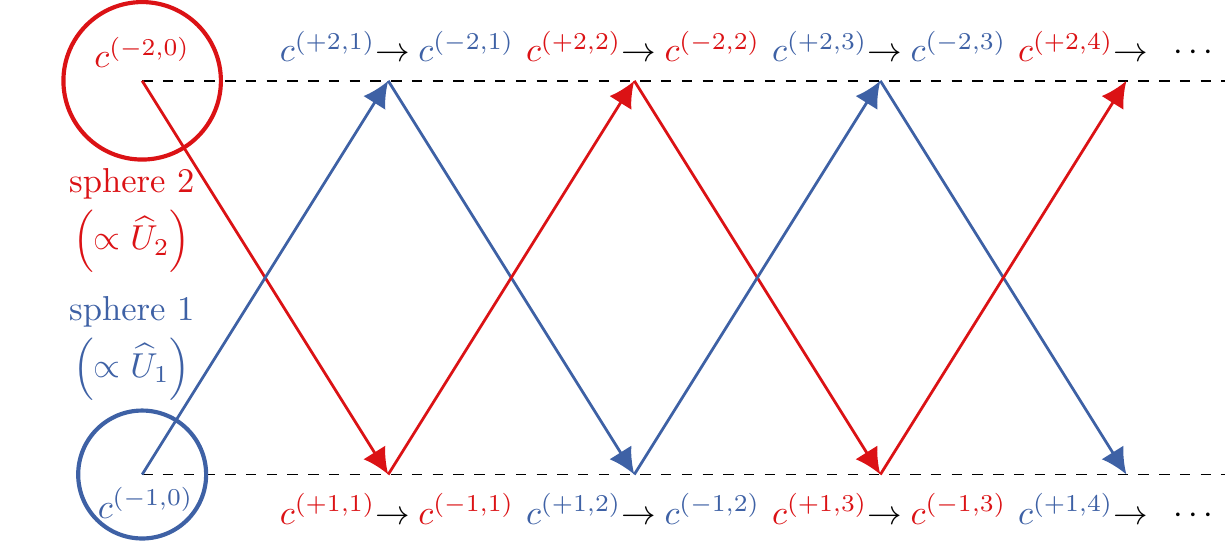}
	\caption{Schematic of the method of reflections for calculating the Brinkman flow field around two spheres undergoing axisymmetric relative motion.
A set of coefficients $c^{(-k,0)}$ represents the exact solution of the flow about sphere $k$ in unbounded fluid, but the presence of this flow violates the boundary conditions at the other sphere $m$.
This incident flow can be described in terms of harmonics about sphere $m$ by a set of coefficients $c^{(+m,1)}$, and the spheres' resistance to deformation creates a reflection flow $c^{(-m,1)}$ which exactly cancels out the incident flow at sphere $m$.
However, this induces a new incident flow at sphere $k$ described by $c^{(+k,2)}$.
This process can be repeated and truncated after a specified number of iterations.
Each set of coefficients are linearly related to the those they are induced by.
Here, all the red coefficients are proportional to $\hU_2$, and all the blue to $\hU_1$.}
	
	\label{fig:reflections}
\end{figure}

The motion of two spheres in a Brinkman theorem, especially in an axisymmetric configuration, has been the subject of much study.
\citet{Kim1985} developed a set of Fax\'en laws relating the force and moment on a sphere in a Brinkman medium to the background flow at its location.
This was accomplished using the well-known method of reflections from low Reynolds number flow theory, applicable in Brinkman media because of the linearity of the Brinkman equations.
An initial approximation to the solution about two spheres is established by considering the Brinkman solution to flow about each sphere in unbounded fluid.
The presence of each sphere's flow does not vanish on the other sphere's surface. Consequently, correction flows must be added to correct this violation of the boundary conditions.
In the limit of infinite reflections, the series of flows converges to the solution. In practice, the sum can be truncated at a desired precision.

\citet{Kim1985} define these correction flows in integral form, a convenient choice for calculating the force and torque on the spheres.
They note that the process can be completed using an explicit velocity representation using addition theorems to describe flows about one sphere around the other using spherical harmonics.
Recently, \citet{liu2020} have employed this approach to devise a solution for Brinkman flow in the presence of two spheres undergoing general relative motion.
Here, we have developed a more limited procedure restricted to axisymmetric motion in the absence of torsion, significantly reducing the space of basis functions.

\subsection{General solution about one sphere}

The general axisymmetric solution to the Brinkman equations can be written in terms of two scalar functions $\chi=\chi(\vx)$ and $\phi=\phi(\vx)$, where the flow fields
\begin{equation}
    \vu = \gr \phi + \bnabla\times\bnabla\times\left(\vx \chi\right), \qquad p = -\alpha^2 \mu \phi \label{eq:Brinkman_gen}
\end{equation}
are a solution to the Brinkman equations if $\phi$ is harmonic and $\chi$ satisfies the Helmholtz equation,
\begin{equation}
    \left(\nabla^2 - \alpha^2\right)\chi=0.
\label{eq:Helmholtz}
\end{equation}
The presence of the Helmholtz equations significantly alters the structure of the flow and the method of solution as compared to the Stokes equations.
First, $\chi = \chi(\alpha r,\theta)$ based on dimensional considerations, suggesting the aforementioned boundary layer effects manifest purely through the function $\chi$.
Second, the Helmholtz equation is not separable in bispherical coordinates, so exact solutions in this coordinate system are not possible, unlike in the case of the Stokes equations.

General solutions $f=f(\vx)$ to each of the scalar functions can be written in terms of the Laplace spherical harmonics $Y_l(\theta)$ defined in \appref{app:harmonics} such that
\begin{equation}
    f(\vx) = \sum_{l=0}^\infty f_l(r) Y_l(\theta), \label{eq:Spherical_Harmonics}
\end{equation}
where the $f_l$ are a series of radially varying functions.
Solutions to the Laplace equation are
\begin{equation}
    \phi_l^- = \frac{1}{r^{l+1}}, \qquad \phi_l^+ = r^l,
\end{equation}
giving rise to the so-called solid harmonics.
The Helmholtz solutions to \eqref{eq:Helmholtz} are
\begin{equation}
    \chi^{-}_l = k_l(\alpha r), \qquad \chi^+_l = g_l(\alpha r),
\end{equation}
where $g_l$ and $k_l$ are the modified spherical Bessel functions of the first and second kind, respectively.
The superscripts $+$ and $-$ denote solutions that are regular and singular as $r\to 0$, respectively.
These four functions provide four basis functions for Brinkman flows at the $l$th mode of spherical harmonics,
\begin{equation}
	\vu_{l0} = \bnabla (\phi_l^- Y_l), \ \vu_{l1} = \bnabla \times \bnabla \times (\vx \chi_l^- Y_l), \ \vu_{l2} = \bnabla (\phi_l^+ Y_l), \ \vu_{l3} =  \bnabla \times \bnabla \times (\vx \chi_l^+ Y_l).
\label{eq:Brinkman_basis}
\end{equation}
In this way the scalar fields may be discretised and represented by a series of coefficients.
The solution can be obtained to desired accuracy by truncating to $n_l$ terms in the spherical harmonic expansion, representing the scalar field via a vector in $\mathbb{R}^{4n_l}$.

We will denote a set of coefficients defined about sphere $k$ with a superscript label, $c^{(k)}$, describing a flow $\vu^{(k)} = \vu^{(-k)} + \vu^{(+k)}$ such that
\begin{equation}
	\vu^{(-k)} = \sum_{l=1}^{n_l} \left(c_{l0}^{(k)} \vu_{l0} + c_{l1}^{(k)} \vu_{l1}\right), \qquad \vu^{(+k)} = \sum_{l=1}^{n_l} \left(c_{l2}^{(k)} \vu_{l2} + c_{l3}^{(k)} \vu_{l3}\right).
\end{equation}
We similarly denote $c^{(+k)}_{lj}$ and $c^{(-k)}_{lj}$ as the regular and singular coefficients of $c^{(k)}_{lj}$, such that for the former $j$ takes the values 0 and 1 and for the latter, 2 and 3.
Note the sum starts from $l=1$ because the $l=0$ flow modes violate the incompressibility condition.
For flows about an isolated sphere in unbounded flow, $\vu^{(+k)}= \vec{0}$, so the flow vanishes as $r \to \infty$.
Note as well that $\vu_{l1}$ is confined to the boundary layer, since it depends only on $\chi_l(\alpha r)$.
Now, we construct the vector of coefficients $c^{(k)}_{lj}$ about sphere $k$ in the presence of another sphere by using the method of reflections.

\subsection{Solution procedure about two spheres}
As shown by \figref{fig:reflections}, we begin by considering a superposition of flows describing motion of each sphere in unbounded flow without regard to the presence of the other.
We will iteratively introduce additional flows (``reflections'') to correct the violation of boundary conditions stemming from existing terms.
Let the coefficients $c^{(-k,j)}$ describe the singular coefficients of the flow about sphere $k$ at the $j$th step in the reflection.
We let $c^{(k,0)}$ be  the ``zeroth''-reflection, i.e.\@ the flow about an isolated sphere, such that $c^{(+k,0)}_{lm} = 0$.
We let $c^{(+k,1)}$ denote the regular modes describing the flow incident at sphere $k$ originating at sphere $m \ne k$.
By linearity, there exists an expression
\begin{equation}
	c^{(+k,j)}_{ln} = A^{mk}_{lnop} c^{(-m,j-1)}_{op},
\end{equation}
relating the two sets of coefficients via some interaction tensor $A$.
Similarly, each singular reflection flow must be linearly related to the incident flow it corrects, i.e.\@
\begin{equation}
	c^{(-k,j)}_{ln} = I^{k}_{lnop} c^{(+k,j)}_{op}
\end{equation}
for some induction tensor $I$.
The functional forms of $A$ and $I$ are reported in \appref{app:harmonics}, and together they comprise a total reflection tensor $R$,
\begin{equation}
    c_{ln}^{(-k,j)} = R^{mk}_{lnop} c^{(-m,j-1)}_{op}, \qquad R^{mk}_{lnop} = I^k_{lnqs} A^{mk}_{pqop}.
\end{equation}
Note the coefficients $c^{(k,0)}$ must be linearly related to the velocity $\hU_k$, but that $c^{(k,1)}$ will be proportional to $\hU_m$ as demonstrated in \figref{fig:reflections}.
In general, $c^{(k,j)}$ contains a factor $\hU_k$ if $j$ is even and $\hU_m$ if $j$ is odd.
Introducing the notation
\begin{equation}
	\hU^{(k,j)} = \begin{cases} \hU_k, & \text{ if } j\text{ mod }2 = 0, \\ 
\hU_m, & \text{ otherwise,} \end{cases}
\end{equation}
we can re-write the coefficients above as $c^{(k,j)} \to \tilde{c}^{(k,j)}$ and let $\tilde{c}^{(k,j)} = \hU^{(j,k)} c^{(j,k)}$ for a set of normalised coefficients $c^{(j,k)}$ which are agnostic to the sphere velocity magnitudes.
The unknown amplitudes $\hU_1$ and $\hU_2$ can be determined by noting, as before, that $\hF_j^{(i)} = C_{jk} \hU_k$ for a matrix $C_{jk}$.
Given $n_r$ reflections, the coefficients of $C$ may be calculated according to
\begin{equation}
    C_{jk} = \begin{cases} iM^2 \left(\frac{4\pi a_k^3}{3} \right) - \sum_{m=0}^{\lfloor n_r/2 \rfloor} c_{1n}^{(j,2m)} F_{nj}, & j=k, \\ 
    -\sum_{m=0}^{\lceil n_r/2 \rceil - 1} c_{1n}^{(j,2m+1)} F_{nj}, &j \ne k, \end{cases} \label{eq:CU}
\end{equation}
where $F_{nk}$ is the force applied to sphere $k$ by the $n$th $l=1$ basis flow harmonic $\vu_{1n}$,
\begin{equation}
    F_{nj} = \int_{\overline{\p\Omega}_j} \vt_{1n} \bcdot \zhat dS,
\end{equation}
and $\vt_{ln}$ is the traction associated with $\vu_{ln}$.
Then $\hU_k$ can be calculated just as in \eqref{eq:unknown_U}.

Note the swim speeds \eqref{eq:swim_speeds} as determined by the reciprocal theorem also depend on the Stokes flow $\vu'$ corresponding to a static dimer being towed through the medium.
This flow can also be represented in terms of a harmonic expansion though the method of reflections, and the process of doing so is similar to the above.
Details can be found in e.g.
\citet{happel2012}. 

\subsection{Asymptotic contributions to swim speed}
{ \color{black}
Using this approach, we construct asymptotic approximations to the swim speed $\bU$ and the contributions $\bU_b + \bU_r = \bU$ related to the steady boundary velocity and Reynolds stress.
Because we are interested in the leading order description of the behaviour, we consider the first two modes' $(n_l = 2)$ interactions through a single reflection $(n_r = 1)$.
In the description of the swim speeds via the reciprocal theorem \eqref{eq:swim_speeds}, each contribution depends quadratically on the Brinkman amplitude fields $\hvu$ and $\hvu^*$.
Since each term in the harmonic expansion of $\hvu$ is proportional to $\hU_1$ or $\hU_2$, each term in the expansions of $\bU_b$ and $\bU_r$ will be proportional to $\hU_1 \hU_2^*$, $|\hU_1|^2$ or $|\hU_2|^2$.
Note that since $\hU_1 = |\hU_1|e^{i(\pi+\phi_1)}$ and $\hU_2 =|\hU_2| e^{-i \phi_2}$ as in \eqref{eq:phi} and \figref{fig:sphere_diagram}(c), then
\begin{equation}
    \hU_1 \hU_2^* = -|\hU_1| |\hU_2| e^{i\phi}, \qquad \phi = \phi_1 + \phi_2,
\end{equation}
where $\phi$ is the phase difference between the spheres' extrema.
This provides a method of categorising the physical origin of each contribution to the swim speed in terms of flow- or geometry-driven interactions, illustrated in \figref{fig:amplitudes}.

\begin{figure}
	\centering
	\includegraphics{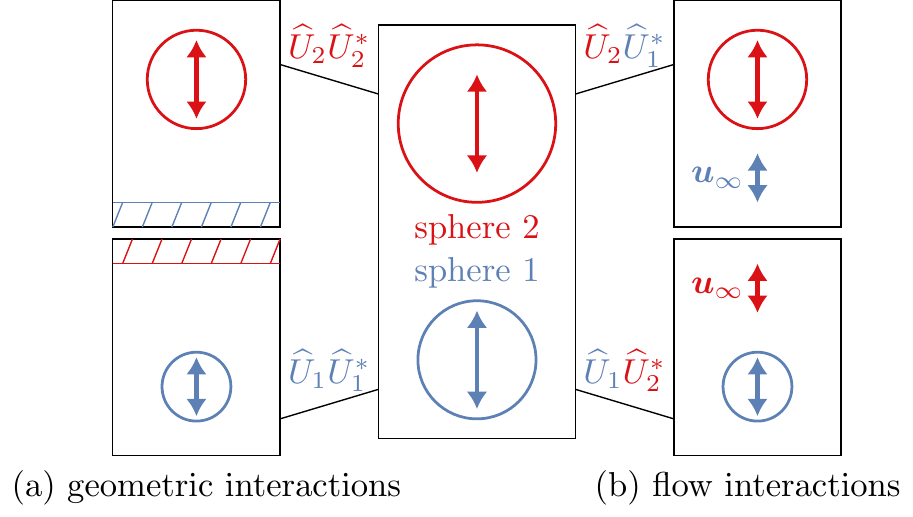}
    \caption{Schematic of types of sphere interactions and their relation to products of the complex Fourier amplitudes $\smash{\hU_1}$ and $\smash{\hU_2}$.
Sphere interactions can be decomposed into geometric and flow contributions.
The presence of the other sphere creates a time-invariant geometric asymmetry akin to an oscillating sphere near a wall (left).
Each sphere is also moving within an oscillating background flow generated by the motion of the other (right).
	By linearity, we can associate products of the velocity amplitudes $\smash{\hU_j \hU_k^*}$ with the different interaction types.
	The geometric interactions correspond to products of each sphere's amplitude with itself, and the flow interactions to the product of both spheres' amplitudes.
	}
	\label{fig:amplitudes}
\end{figure}

Contributions to the speeds proportional to $\hU_1 \hU_2^*$ correspond to interactions between one sphere's motion and the oscillating background flow induced by the other.
In the presence of a nonzero phase difference $\phi \ne 0$, this interaction is not time-reversible, and thus can give rise to motion even in the absence of fluid inertia ($M^2 = 0$), since the scallop theorem does not apply.
We refer to these terms in the swim speeds as ``flow interactions.''
On the other hand, contributions proportional to $|\hU_1|^2$ and $|\hU_2|^2$ arise from the effects on one sphere's motion caused by the other's presence, which breaks geometric symmetry.
We refer to these as ``geometric interactions.''
}

\subsubsection{Velocity magnitudes and phase difference}

{\color{black} 

Recall from \eqref{eq:CU} that we can relate the fluid force amplitude $\smash{\hF_k^{(f)}}$ with the two sphere amplitudes $\smash{\hU_k}$ via a mobility matrix defined by the harmonic expansion coefficients. 
This is sufficient to find the asymptotic dependence of the velocity amplitudes on the fluid inertia $M^2$ (reported in \tabref{tab:vel_mags}, columns two and three) or, for the Stokes case, the solid inertia $S^2$ (reported in \tabref{tab:vel_mags}, columns four and five.)
In the Stokes limit, we assume the dimensionless ratio $\rho_2/\rho$ becomes large such that $S^2 = 2M^2\rho_2/3\rho$ takes a finite value in the limit $M^2 \to 0$.
We introduce a rescaling of the actual mass of sphere $k$,
\begin{equation}
    \tilde{m}_k = a_k^3 \frac{\rho_k}{\rho_2} \label{eq:eff_mass_alt}
\end{equation}
such that $2 M^2 m_k = 3 S^2 \tilde{m_k}$.
The Stokes scalings in \tabref{tab:vel_mags} are reported in terms of this value.

\begin{table}
    \centering
    {\tabulinesep=1.2mm
    \begin{tabu}{c|c|c|c|c}
        quantity & $ 0 < M^2 \ll 1 $ & $1 \ll M^2$ & $0 < S^2 \ll 1$ & $ 1 \ll S^2$ \\[0.25em]
        $|\hU_1|$ & $\dfrac{a_2}{a_1+a_2}$ & $\dfrac{m_2}{m_1+m_2}$ &  $\dfrac{a_2}{a_1+a_2}$ & $\dfrac{m_2}{m_1+m_2}$ \\ 
        $|\hU_2|$ & $\dfrac{a_1}{a_1+a_2}$ & $\dfrac{m_1}{m_1+m_2}$ & $\dfrac{a_1}{a_1+a_2}$ & $\dfrac{\tilde{m}_1}{\tilde{m}_1+\tilde{m}_2}$\\ 
        $\sin \phi$ & $\dfrac38(a_2 - a_1) d M^2$ & $\dfrac{9}{2}\dfrac{a_1^2 m_2 - a_2^2 m_1}{\sqrt{2} \ m_1 m_2} \dfrac{1}{M}$ & $\dfrac29 \dfrac{a_1 \tilde{m}_2 - a_2 \tilde{m}_1}{a_1 a_2} S^2$ & $\dfrac{9}{2}\dfrac{a_1 \tilde{m}_2 - a_2 \tilde{m}_2}{\tilde{m}_1 \tilde{m}_2} \dfrac{1}{S^2}$\\
        $\cos \phi$ & 1 & 1 & 1 & 1
    \end{tabu}}
    \caption{Leading-order scalings of the velocity amplitude magnitudes $|\hU_k|$ and trigonometric functions of the sphere phase difference $\phi$. The second and third columns correspond to the small- and large-inertia limits in the presence of nonzero fluid inertia $(M^2 \ne 0)$, and the fourth and fifth columns are the same limits for Stokes flow $(M^2 = 0)$. For the $M^2 \ne 0$ case, values are given in terms of the effective mass $m_k$ \eqref{eq:eff_mass}. For Stokes flow, rescalings of the actual sphere masses $\tilde{m}_k$ \eqref{eq:eff_mass_alt} are used.}
    \label{tab:vel_mags}
\end{table}

In both cases, the velocity magnitudes undergo a transition from depending on the sphere sizes at low inertia levels to sphere masses at high inertia levels, and the phase difference $\phi$ vanishes in both limits as the sphere--sphere interactions become dominated by Stokes drag or added mass effects, respectively.
However, the rate at which the phase difference changes is not the same, as observed in the functional form of $\sin \phi$.
In the fluid inertial case, we observe that at low inertia, the larger sphere leads the larger sphere since $\phi > 0$ for $a_2 > a_1$ and $\sin \phi \propto (d M)[(a_2-a_1)M]$.
This factor of $d$ in the proportionality relation stems from the reduction of the Brinkman boundary layer width as $M$ increases.
At high inertia, there is a more complicated comparison for determining the leading sphere which involves both size and mass asymmetries, as $\phi > 0$ if $a_1^2 m_2 > a_2^2 m_1$.
Due to the presence of the Basset force which is proportional to $M$, the phase difference falls off as $1/M$. In the Stokes case, there is no boundary layer evolution and no Basset force.
The relevant asymmetry is the same in both low and high inertial limits, as $\phi > 0$ if $a_1 \tilde{m}_2 > a_2 \tilde{m}_1$, and the dependence on $S$ in the proportionalities is always quadratic since the added mass effect is the only inertial coupling between the spheres.

Below, we report the leading-order contributions for each combination of inertial limit ($M^2 \ll 1$ vs.\@ $M^2 \gg 1$), mechanism (boundary velocity vs.\@ Reynolds stress), and interaction type (flow vs.\@ geometric). See \appref{app:speed_calcs} for the complete derivation. For ease of notation we introduce the quantities
\newcommand{\mi}[1]{#1_{\smash{\scalebox{0.8}{\hspace{1pt}-}}}}
\newcommand{\pl}[1]{#1_{\smash{\scalebox{0.6}{+}}}}
\newcommand{\di}[1]{\left[#1\right]}
\begin{equation}
    \pl{a} = a_1 + a_2, \qquad \mi{a} = \frac{1}{a_1^{-1} + a_2^{-1}}, \qquad \di{a}  = \frac{a_2 - a_1}{a_1 + a_2}, \nonumber
\end{equation}
\begin{equation}
    \pl{m} = m_1 + m_2, \qquad \mi{m} = \frac{1}{m_1^{-1} + m_2^{-1}}, \qquad \di{m} = \frac{m_2 - m_1}{m_1 + m_2}.
\end{equation}
The quantities $\pl{a}$ and $\mi{a}$ approach the large and small sphere sizes as the difference between them becomes large, and $\di{a}$ is a dimensionless measure of the size asymmetry.
The $m$ symbols are the same, with respect to the effective sphere masses $m_k$.

\subsubsection{Stokes flow}

In the Stokes limit ($M^2 = 0$), the leading-order boundary velocity contribution is
\begin{equation}
\bU_b = -\frac{3\mi{a}}{2d^2} |\hU_1| |\hU_2| \sin \phi,
\end{equation}
consistent with previous investigations of inertial dimers in Stokes flow \citep{hubert2021}.
Note that $\phi \ne 0$ is required for motion according to the scallop theorem, that this symmetry is broken by the presence of solid inertia ($S^2 > 0$), and that the dimer swims towards the leading sphere (i.e. towards sphere 1 when $\phi > 0$.)
As mentioned, the sign of $\sin \phi$ and thus $\bU_b$ is the same in both limits, since it depends in both cases on the quantity $a_1 \tilde{m}_2 - a_2 \tilde{m}_1.$
By definition \eqref{eq:swim_speeds} there is no Reynolds stress contribution in the Stokes case, so $\bU = \bU_b$. Substituting in the values for $|\hU_k|$  and $\phi$ in \tabref{tab:vel_mags} yields
\begin{equation}
    \bU \approx \begin{cases} \dfrac{\mi{a}}{3 \pl{a}^2 d^2}\left(a_2 m_1 - a_1 m_2\right) S^2, & 0 < S^2 \ll 1 \\[0.5em] \dfrac{27 \mi{a}}{4\pl{m}^2 d^2}\left(a_2 m_1 - a_1 m_2\right) \frac{1}{S^2}, & S^2 \gg 1. \end{cases}
\end{equation}
In the case of $\rho_1 = \rho_2$, this implies motion towards the small sphere as observed in our numerical results and the work of \citet{felderhof2016} and \citet{hubert2021}.

\subsubsection{Mechanisms at low inertia}
For nonzero fluid inertia $0 < M^2 \ll 1$, the leading-order boundary velocity contribution is
\begin{multline}
    \bU_b = \left[-\frac{3\mi{a}}{2 d^2} \sin\phi + \frac{3 \pl{a} \mi{a}}{4\sqrt{2} d^2} \di{a} M\cos \phi\right]|\hU_1||\hU_2| \\ + \frac{3 \mi{a}}{4\sqrt{2}} M\left(a_2 |\hU_2|^2 - a_1 |\hU_1|^2\right). \label{eq:Ub_low_first}
\end{multline}
It may appear that the Stokesian term will be small compared to the others as $M$ grows.
However, substituting the expressions in \tabref{tab:vel_mags} for the velocity magnitudes shows the $\cos\phi$ part of the flow interaction cancels the geometric interaction at leading order,
\begin{equation}
    \bU_b = -\frac{3\mi{a}^2}{2 \pl{a} d^2} \sin \phi - \frac{3\mi{a}^2}{2\sqrt{2} d^2} \di{a}  M \sin^2\left(\frac{\phi}{2}\right),
\end{equation}
so that as before the dimer swims towards the leading sphere.
Upon substitution for $\sin \phi$, the first term dominates and
\begin{equation}
    \bU_b = \mi{C^{(b)}} M^2, \qquad \mi{C^{(b)}} =  -\frac{9\mi{a}^2}{16 d} \di{a}. \label{eq:Ub_low}
\end{equation}
The boundary velocity contribution always promotes motion towards the smaller sphere, proportional to the dimensionless size asymmetry $\di{a}$.
Boundary layer effects are manifested in the scaling for $\sin \phi \propto d M$, so that $\mi{C^{(b)}} \propto 1/d$ rather than $1/d^2$ as in the other coefficients arising from the the method of reflections in \eqref{eq:Ub_low_first}.
Numerical results for $a_1 \in [0.1,0.9]$ and $d \in \{3,5,10\}$ are shown in the bottom row of \figref{fig:scalings_low}, normalised by the proportionality in \eqref{eq:Ub_low}.
The agreement is good for $Md < 1$, when each sphere lies within the Brinkman boundary layer surrounding the other.

\begin{figure}
	\centering
	\includegraphics{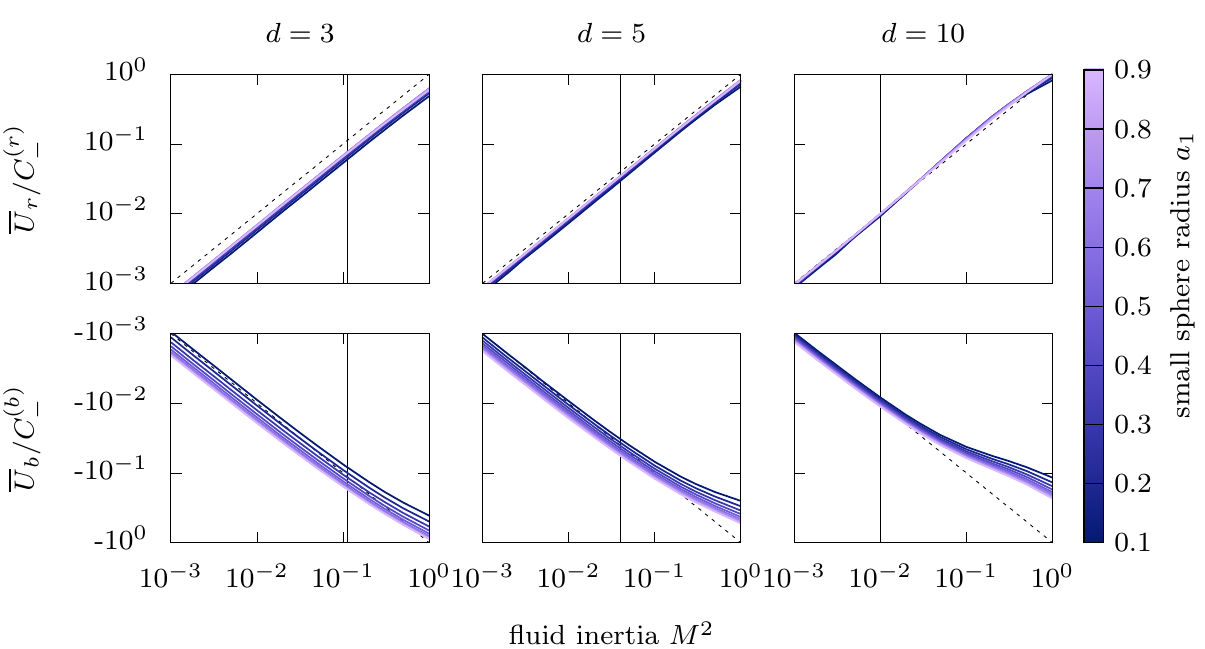}
	\caption{Normalised swim speeds at low fluid inertia for an equal-density swimmer with $\rho_1 = \rho_2 = 1$.
	The swim speed contributions from the boundary velocity $\smash{\bU_b}$ (bottom) and Reynolds stress $\smash{\bU_r}$ (top) are plotted with respect to $M^2 \le 1$ for a variety of small sphere radii $a_1$ (denoted by line colour) and separation distances $d$ (separated by column.)
 In each plot, a black vertical line marks where $M d = 1$.
 For points to the left of this line, the spheres lie within the others' Brinkman boundary layer.
 To the right of the line, the boundary layers are small and do not extend to the other sphere.
	The speeds are scaled by $\smash{C^{(b/r)}_{-}}$, given by \eqref{eq:Ub_low} and \eqref{eq:Ur_low}, and lie parallel to the dashed line of slope 1, showing that in this region $\smash{\bU_{b/r} \approx C^{(b)}_- M^2}$ within the boundary layers.}
	\label{fig:scalings_low}
\end{figure}

The leading-order Reynolds stress contribution takes the form
\begin{equation}
    \bU_r = \frac{15 \pl{a}^2 \mi{a}}{32 d^2} \di{a} M^2 |\hU_1| |\hU_2|  \cos \phi.
\end{equation}
There is no contribution from geometric interactions here because it can be shown to vanish at order $M^2/d^2$ (see \appref{app:speed_calcs}.)
Substitution of the values in \tabref{tab:vel_mags} yields
\begin{equation}
    \bU_r = \mi{C^{(r)}} M^2, \qquad
    \mi{C^{(r)}} = \frac{15 \mi{a}^2 \pl{a} }{32 d^2} \di{a} M^2, \label{eq:Ur_low}
\end{equation}
so the Reynolds stress contribution always promotes motion towards the larger sphere.
Numerical comparisons to the scaling are shown in the top row of \figref{fig:scalings_low}.
As in the case of the boundary velocity contribution, there is good agreement with the computed scaling, although there is less variability with respect to both the small sphere radius and separation distance $d$.
As in that case, however, the agreement is better for $Md < 1$.

\subsubsection{Mechanisms at high inertia}

In the large inertia limit $M^2 \gg 1$, the boundary velocity takes the form
\begin{multline}
    \bU_b = \left[\frac{1}{2\sqrt{2}} \frac{\pl{a}^2 \mi{a}^2 M}{ d^4} \left(\di{a} \cos \phi + \sin \phi\right) \right]|\hU_1||\hU_2| \\ + \frac{3}{4\sqrt{2}} \frac{\mi{a} M}{d^2} \left(a_2 |\hU_2|^2 - a_1 |\hU_1|^2\right).
\end{multline}
Since $Md > 1$ in this limit, the spheres do not exist within each other's boundary layers, and as a result the flow interactions are weaker $(\propto 1/d^4)$ than in the small-inertial limit $(\propto 1/d^2)$.
The geometric interactions remain the same strength $(\propto 1/d^2)$ and thus dominate at order $M$.
Substituting in the values in \tabref{tab:vel_mags} gives
\begin{equation}
\bU_b \approx C_+ M, \qquad C_+ = \frac{3}{4\sqrt{2}} \frac{\mi{a}}{d^2 \pl{m}^2} \left(a_2 m_1^2 - a_1 m_2^2\right). \label{eq:Ub_high}
\end{equation}

Performing the same analysis for the Reynolds stress shows that at order $M$, the flow and geometric effects exactly cancel.
However, there exists an additional geometric contribution at order 1 which is not present in the boundary velocity contribution,
\begin{multline}
    \bU_r = -\left[\frac{1}{2\sqrt{2}} \frac{\pl{a}^2 \mi{a}^2 M}{ d^4} \left(\di{a} \cos \phi + \sin \phi\right) \right]|\hU_1||\hU_2| \\ - \frac{3}{4\sqrt{2}} \frac{\mi{a} M}{d^2} \left(a_2 |\hU_2|^2 - a_1 |\hU_1|^2\right) + \frac{9\mi{a}}{8d^2}\left(|\hU_1|^2 - |\hU_2|^2\right). \label{eq:Ur_high_first}
\end{multline}
At leading order, then, the isolated Reynolds stress contribution is
\begin{equation}
    \bU_r \approx -C_+ M. \label{eq:Ur_high}
\end{equation}
Numerical results comparing to the scalings \eqref{eq:Ub_high} and \eqref{eq:Ur_high} are shown in \figref{fig:scalings_high}.
In contrast to the low-inertia case, there is little variation with respect to $a_1$ or $d$.
The agreement with the scalings is good, especially at higher $M$ where the effects of the constant term in \eqref{eq:Ur_high} become negligible.

\begin{figure}
	\centering
	\includegraphics{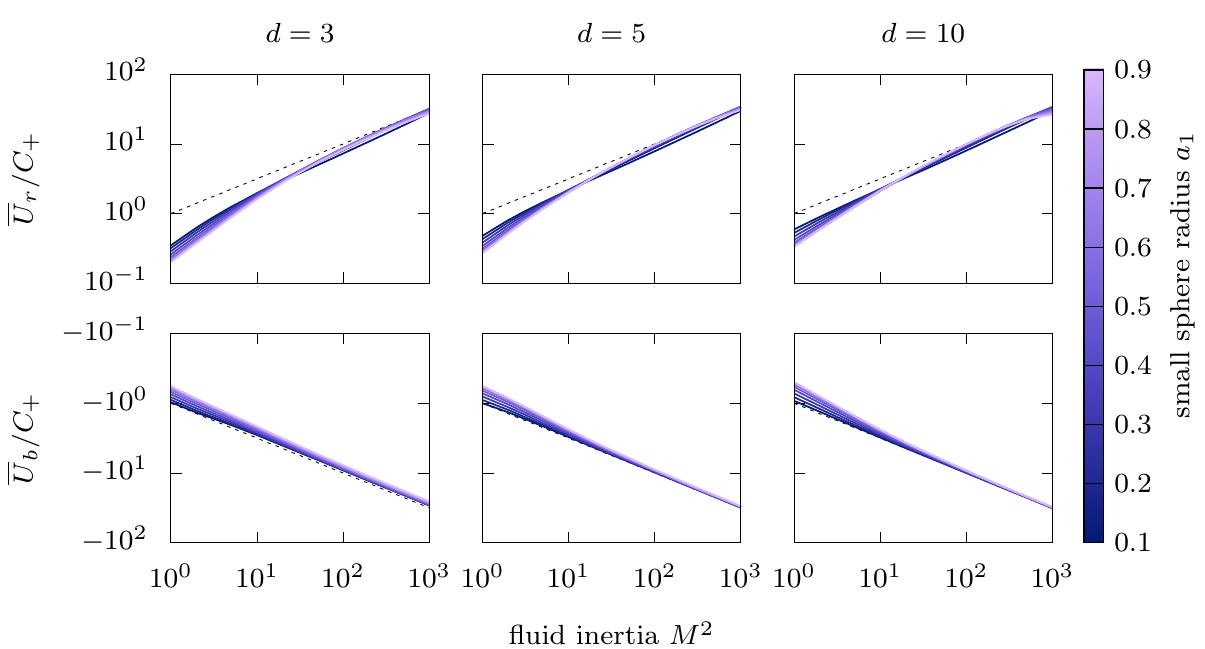}
	\caption{
	Normalised swim speeds at high fluid inertia for an equal-density swimmer with $\rho_1 = \rho_2 = 1$.
	The swim speed contributions from the boundary velocity $\smash{\bU_b}$ (bottom) and Reynolds stress $\smash{\bU_r}$ (top) are plotted with respect to $M^2 \ge 1$ for a variety of small sphere radii $a_1$ (denoted by line colour) and separation distances $d$ (separated by column.)
	The speeds are scaled by the large--$M^2$ coefficient $\smash{C_+}$ from \eqref{eq:Ub_high} and \eqref{eq:Ur_high}.
The lines' approaching the 1/2 slope of the dashed line indicate that at high levels of inertia \smash{$\bU_{b/r} \approx C_+^{(b/r)}M$}.
There is little variation with respect to $d$, as the spheres exist outside of each others' boundary layers for all of the provided inertial range.
	}
	\label{fig:scalings_high}
\end{figure}

\subsubsection{Total swim speed}
Using the scalings for each mechanism listen in the previous subsection, we can determine the asymptotic form of the total swim speed in the low- and high-inertial limits.
At $0 < M^2 \ll 1$, the boundary velocity scaling \eqref{eq:Ub_low}, which scales as $1/d$, dominates that of the Reynolds stress \eqref{eq:Ur_low}, at $1/d^2$. As such,
\begin{equation}
    \bU \approx -\frac{9 \mi{a}^2}{16 d} \di{a} M^2,
\end{equation}
so that across the entire parameter space, we expect motion towards the small sphere at small-but-finite levels of fluid inertia.

At $M^2 \gg 1$, the total swim speed $\bU = \bU_b + \bU_r$ will have no order-$M$ contribution, since the $C_+ M$ terms in \eqref{eq:Ub_high} and \eqref{eq:Ur_high} will cancel. We are left with the order-1 term in \eqref{eq:Ur_high_first}, which gives a theoretical limiting speed as $M^2 \to \infty$,
\begin{equation}
    \bU \approx \frac{9 \mi{a}}{8 d^2} \di{m} =: \bU_\infty, \label{eq:U_inf}
\end{equation}
showing that at high inertia motion is always directed towards the more massive sphere.
This confirms the impression formed by \figref{fig:first_speeds}(a--b) that at $M^2 \ll 1$ motion is directed towards the smaller sphere and at $M^2 \gg 1$ towards the heaver sphere.

Numerically calculated swim speeds for several dimers with $a_1 = 1/2$ and $d=3$ (the same as in \figref{fig:steady_flows} and \figref{fig:first_speeds}) are plotted in \figref{fig:U_inf}.
Each line corresponds to a different effective mass ratio $m_1/m_2$, and as before the densities are chosen such that the dimer is neutrally buoyant.
In panel (a), the raw speeds are plotted as a function of inertia, yielding a range of velocities at the highest value, $M^2 \approx 1000$.
In panel (b), the speeds are normalised by the limiting factor $\bU_\infty$, collapsing for $M^2 \gg 1$ onto a single trajectory.

\begin{figure}
    \centering
    \includegraphics{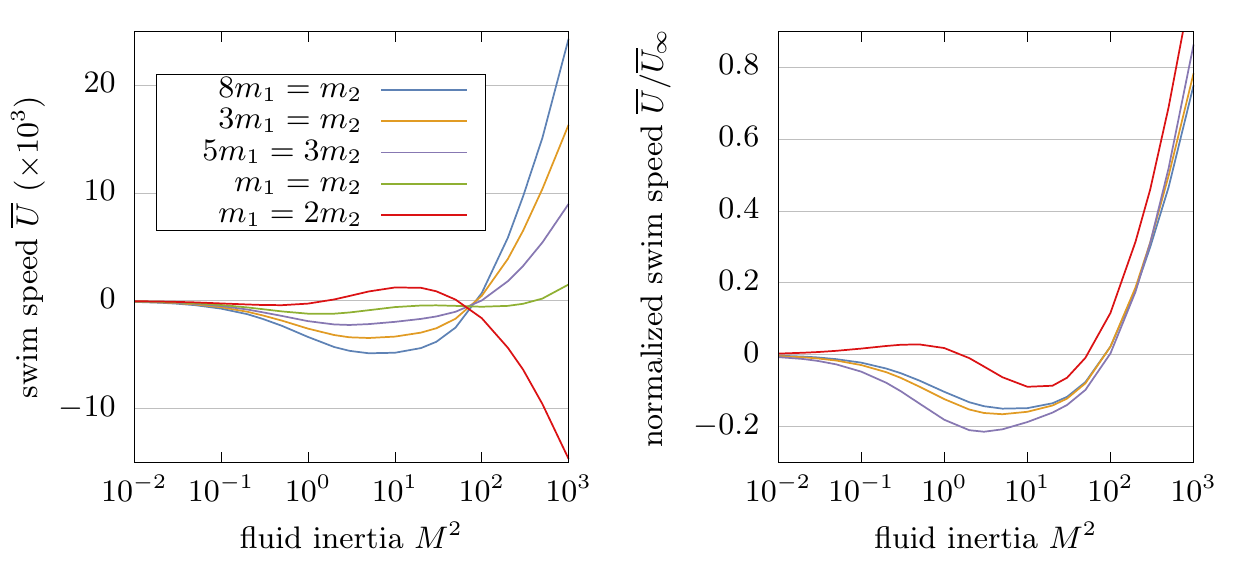}
    \caption{Plots of the total swim speed $\bU$ as a function of fluid inertia $M^2$. (a) The swim speed is plotted for a system with $a_1 = 1/2$ and $d = 3$ for a variety of effective mass ratios $m_1/m_2$. Positive values correspond to motion near sphere 2. At high $M^2$, motion is towards the more massive sphere. (b) The normalised swim speed $\smash{\bU / \bU_\infty}$ is plotted over the same range of inertia. Positive values here correspond to motion in the same direction as the theoretical limiting value $\bU_\infty$ \eqref{eq:U_inf} as $M^2 \to \infty$. The collapse indicates consistent scaling across mass ratios. At $M^2 \approx 1000$, near the upper limit of the intermediate $\Rey$ range, each of the systems' swim speeds is still growing with increased inertia, reaching about 80\% of the theoretical limit.}
    \label{fig:U_inf}
\end{figure}

For the single sphere case, \citet{felderhof2017} derived analogous asymptotic limits in terms of the multipole coefficients of the swim stroke.
There are a few differences worth noting which stem from the dimer geometry.
First, for a single sphere, the boundary is always subject to the Brinkman boundary layer.
Thus, the introduction of the separation distance $d$ represents a unique aspect of the parameter system.
In particular, it yields two separate transitions.
The low-$M^2$ scalings in \figref{fig:scalings_low} are valid only until $M d > 1$ and the spheres exit each others' boundary layers.
However, the high-$M^2$ scalings in \figref{fig:scalings_high} do not show good agreement until $M > 1$, when the added mass effect dominates the velocity amplitude ratios in \tabref{tab:vel_mags}. 
These multiple transitions give rise to the double-direction switch observed in \figref{fig:first_speeds}(b).

Secondly, while multipole coefficients are widely useful in a wide variety of contexts including squirmer analysis \citep{Pedley16}, the dimer geometry provides a clear and intuitive understanding of the results in terms of different types of asymmetries in the problem.
We clearly see the transition from a dependence on size asymmetry (due to the domination of Stokes drag) to mass asymmetry (due to the domination of the added mass effect.)
One difference from the one-sphere analysis is that \citet{felderhof2021} found swimming could proceed in either direction as a function of the surface deformation stroke.
However, in this case the gait (corresponding to the choice of $\hU_1$ and $\hU_2$) is initially unknown and must be solved for as a function of the level of fluid and solid inertia.

}

\section{Conclusions}
In this work, we have investigated the motion of an asymmetric dimer at intermediate Reynolds number.
The distance between the dimer's constituent spheres is set to oscillate with respect to time, as though driven by an internal force exerted by a connecting rod.
While this actuation would yield no motion at $\Rey=0$ according to the scallop theorem, the dimer swims in the presence of inertia.
Recent works have investigated similar systems, in one case restricting to the particle inertial effects of very dense spheres in Stokes flow \citep{gonzalez-rodriguez2009,felderhof2016,hubert2021}, and in the other including the effects of fluid and particle inertia \citep{dombrowski2019}.
We generalise these analyses to include the effects of both fluid and particle inertia in terms of a variant of the Reynolds number $M^2$, thus explaining and unifying the above results.
We also find novel behaviour of the dimer reminiscent of the double direction switches observed by \citet{Collis2017}.

Our calculation of the dimer's swim speed shows the steady flow field is driven by an effective slip velocity and Reynolds stress which can be understood as two mechanisms giving rise to motion.
Each mechanism is driven by two classes of sphere--sphere interactions, between one sphere's motion and 1) the oscillating background flow induced by the other's motion, and 2) a geometric asymmetry induced by the other's presence.
The previous investigations into dense swimmers in Stokes flow correspond only to the flow--flow interaction, since time-invariant geometric asymmetries are not sufficient to evade the effects of the scallop theorem.
Under these conditions, the swim speed of the dense dimer in Stokes flow can be shown as in \citep{hubert2021} to depend in a simple way on the phase difference $\phi$ between the spheres' oscillations.
The speed vanishes as $S^2 \to 0$ and $S^2 \to \infty$, and asymptotic analysis shows the same direction of motion, which is towards sphere 2 if $a_2 \tilde{m}_1 > a_1 \tilde{m}_2$. If the sphere densities are equal, this is towards the small sphere.



In the presence of fluid inertia, the interplay between the four mechanism--interaction combinations yields a richer set of behaviours.
The flow interaction is the primary driver of each mechanism at small $M^2$, while the geometric interaction dominates at large $M^2$.
For $M^2 \ll 1$, each mechanism drives translation of the dimer in opposing directions as observed in \figref{fig:first_speeds}(a.ii) and the scalings \eqref{eq:Ub_low} and \eqref{eq:Ur_low}: the slip velocity causes swimming in the direction of the smaller size sphere, and the Reynolds stress towards the larger.
The contribution to swimming of the slip velocity $(\sim 1/d)$ dominates that of the Reynolds stress $(\sim 1/d^2)$, consistently driving overall motion towards the small sphere.
At larger $M^2 \gg 1$, the leading-order contributions of the two methods cancel as shown in \eqref{eq:Ub_high} and \eqref{eq:Ur_high}, but there is an $\bigO{1}$ part of the swim speed originating from the geometric interaction which drives motion towards the more massive sphere \eqref{eq:U_inf}.

\textcolor{black}{
A similar decomposition was found in the work of \citet{felderhof1994inertial,felderhof2017,felderhof2019effect}.
That analysis also showed boundary velocity effects dominating at low inertia and a balance between boundary velocity and Reynolds stress effects at high inertia, suggesting the physical mechanisms driving variation of the dimer's swim speed in inertial fluid are the same as in the swimming of a deforming single single sphere.
However, the dimer geometry provides an understanding of how symmetry breaking arises out of the spheres' size and mass asymmetry as a function of inertia, as opposed to ascribing it to particular modes of surface deformation.
It also gives rise to an additional transition in the mechanism as $M^2$ grows, stemming from the introduction of the second length scale $d$, the sphere separation distance, which has no analogue in the single-sphere system.}

\textcolor{black}{
When $Md$ becomes larger than 1, a boundary layer around the spheres becomes smaller than their separating distance.
This weakens the flow-mediated interactions and renders the corresponding scalings \eqref{eq:Ub_low} inaccurate.
We may understand this as the transition from the ``low inertia'' to ``high inertia'' parts of Section 4.
Secondly, when $M$ is much smaller than 1, the spheres' relative oscillation is a function of size asymmetry, as the forces they experience are dominated by Stokes drag.
When $M$ becomes larger than 1, this transitions to oscillation as a function of mass asymmetry, because the forces they experience become dominated by the added mass effect.
This represents a switch from the $0 < M \ll 1$ column to the $M \gg 1$ column in the table of velocity and phase scalings \tabref{tab:vel_mags}.
These two transitions are what give ride to the double direction switch observed in \figref{fig:first_speeds}(b) and the work of \citet{Collis2017}.}
\\



\subsection*{Acknowledgements}
N.D.\@ acknowledges support from the National Defense Science and Engineering Graduate Fellowship Program and the Department of Defense, and from the NSF-Simons Center for Mathematical and Statistical Analysis of Biology at Harvard, award number 1764269.
D.K. and T.D acknowledge the National Science Foundation, grant award DMR-1753148. 

\subsection*{Declaration of interests}
The authors report no conflict of interest.

\appendix
\section{Derivation of ans\"atz}\label{app:ansatz}
In this section, we show the ans\"atz \eqref{eq:ansatz} corresponds to the leading terms in a series expansion of the flow fields $(\vu,p)$, which satisfy the dimensionless Navier--Stokes equations \eqref{eq:Navier-Stokes}.
The sphere velocities \eqref{eq:U_k} are $2\pi$-periodic and scale as $U_k \sim \varepsilon$, suggesting the leading-order flow is of order $\varepsilon$ and also $2\pi$-periodic in time.

We introduce the series expansions
\begin{equation}
    \vu = \varepsilon \vu_1 + \varepsilon^2 \vu_2 + \cdots, \qquad p = \varepsilon p_1 + \varepsilon^2 p_2 + \cdots.
\label{eq:series}
\end{equation}
Substituting into \eqref{eq:Navier-Stokes} yields the following equations at the first two orders of $\varepsilon$:
\begin{equation}
    M^2 \frac{\p \vu_1}{\p t} = -\gr p_1 + \nabla^2 \vu_1, \label{eq:order_1}
\end{equation}
\begin{equation}
    M^2 \frac{\p \vu_2}{\p t} = -\gr p_2 + \nabla^2 \vu_2 - M^2 \vu_1 \bcdot \gr \vu_1.
\label{eq:order_2}
\end{equation}
More generally, the substitution transforms the non-linear Navier--Stokes equations into a series of coupled linear PDEs for the order $\varepsilon^k$ flow $\left(\vu_k,p_k\right)$.
Each of these flows can depend non-linearly on the order $\varepsilon^j$ velocity field $\vu_j$, with $j < k$, allowing for successive evaluation and substitution.

Because the order $\varepsilon$ flow is driven by the spheres' periodic motion, we let $\vu_1 = \hvu e^{it}$ and $p_1 = \hp e^{it}$.
Substituting into \eqref{eq:order_1} yields
\begin{equation}
    \left(\nabla^2 - iM^2 \right)\hvu = \gr \hp,
\end{equation}
which is \eqref{eq:Brinkman}.
Now, the quadratic forcing term in \eqref{eq:order_2} can be calculated in terms of $\hvu$, taking the form
\textcolor{black}{
\begin{align}
	-M^2\vu_1 \bcdot \gr \vu_1 &= -M^2 \real\{\vu_1\} \bcdot \real\{\gr \vu_1\} \nonumber \\
	&= -M^2 \left(\frac{\hvu e^{it} + \hvu^* e^{-it}}{2}\right) \bcdot \left(\frac{\gr \hvu e^{it} + \gr \hvu^* e^{-it}}{2}\right) \nonumber \\
	&= -M^2 \left[\left(\frac{\hvu \bcdot \gr \hvu^* + \hvu^* \bcdot \gr \hvu}{4}\right) + \left(\frac{\hvu \bcdot \gr \hvu e^{2it} + \hvu^* \bcdot \gr \hvu^* e^{-2it}}{4}\right)\right] \nonumber \\
	&= -M^2 \left[\frac{\real\{\hvu \bcdot \gr \hvu^*\}}{2} +  \frac{\real\{\hvu \bcdot \gr \hvu e^{2it}\}}{2}\right] \nonumber \\
	&= -\frac{M^2}{2}\hvu \bcdot \gr \hvu^* - \frac{M^2}{2}\hvu \bcdot \gr \hvu e^{i2t} \nonumber \\
	&=: \bvf + \hhvf e^{i2t},
\end{align}
}
\hspace{-0.75em} where we have introduced steady and oscillatory body forces $\bvf$ and $\hhvf e^{i2t}$.
This implies the existence of steady and oscillatory parts of $\vu_2 = \bvu + \widehat{\hvu} e^{i2t}$ which, upon substituting into \eqref{eq:order_2}, yields
\begin{equation}
    0 = \left[\nabla^2 \bvu - \gr \bp +\bvf\right] + \left[\left(\nabla^2 - i2M^2\right) \widehat{\hvu} - \gr \widehat{\hp} + \hhvf\right] e^{i2t}.
\end{equation}
Both bracketed relations must be zero at all times.
Setting the first to zero reproduces \eqref{eq:Stokes}, which describes the steady flow of interest.
The second describes a $\pi$-periodic flow, but the steady contribution resulting from this flow is $\bigO{\varepsilon^4}$, from the advective term $\vu_2 \bcdot \bnabla \vu_2$ present in the order-$\varepsilon^4$ unsteady Stokes equation.
\textcolor{black}{At order-$\varepsilon^2$, the swim speed does not depend on $\widehat{\hvu}$.
Thus, despite being order-$\varepsilon^2$ itself, we omit it from our analysis.
This is consistent with similar treatments in related work \citep{felderhof2017}.}
Truncating the series expansion \eqref{eq:series} after two terms and neglecting the $\pi$-periodic part of the $\varepsilon^2$ term yields the ans\"atz \eqref{eq:ansatz} used in the main text.
As written in the main text, we also adopt the form of the ans\"atz for other variables in the problem.
In particular, we let $p = \varepsilon \hp e^{it} + \varepsilon^2 \bp$,  $U_k(t) = \varepsilon \hU_k e^{it}$, and $U = \varepsilon^2 \bU$.
Following from this is the net force of sphere $k$, $F_k^{(n)} = [i \hU_k \rho_k (4\pi a_k^3/3) ] e^{it} =: \hF_k^{(n)} e^{it}$.

While the boundary conditions for $\vu$ \eqref{eq:bcs} are defined with respect to the moving sphere surfaces $\p\Omega_k$, we can relate them to conditions on $\hvu$ and $\bvu$ on the time-averaged boundaries $\overline{\p\Omega}_k = \{\vx : |\vx - \bz_k \zhat| = a_k\}$.
We let a pair of positions $\vx_k$ denote a position $\vx_k \in \p\Omega_k$ and $\vx_{k,0} \in \overline{\p\Omega}_k$ denote the corresponding position on the time-averaged surface.
The time-dependent displacement between the two points is $\vx_k - \vx_{k,0} = -i\varepsilon \hU_k e^{it} \zhat$, so arbitrary fields matching our ans\"atz $\smash{\psi(\vx) = \varepsilon \widehat{\psi}e^{it} + \varepsilon^2 \overline{\psi}}$ can be evaluated on the moving boundary as
\begin{equation}
	\psi(\vx_k) = \varepsilon \widehat{\psi}(\vx_{k,0})e^{it} + \varepsilon^2 \left[\overline{\psi} - \frac{i\hU_k}{2} \frac{\p \widehat{\psi}^*}{\p z}\right]_{\vx = \vx_{k,0}} + \bigO{\varepsilon^2 e^{i2t}} + \bigO{\varepsilon^3}.
\label{eq:general_bc}
\end{equation}
The factor of 1/2 comes from multiplying the real parts of complex functions, so that the Taylor series term
\begin{align}
	\real\left\{	(\vx_k - \vx_{k,0})\right\} \bcdot \real\left\{\gr \varepsilon \widehat{\psi}\right\} &=\frac{\varepsilon^2}4 \left(-i \hU_k e^{it} + i \hU_k^* e^{-it}\right) \left(\frac{\p\widehat{\psi}}{\p z}e^{it} + \frac{\p \widehat{\psi}^*}{\p z}e^{-it}\right) \nonumber \\
	&= \frac{\varepsilon^2}2 \left[\real\left\{-i \hU_k \frac{\p \widehat{\psi}^*}{\p z} \right\} + \real\left\{-i \hU_k \frac{\p \widehat{\psi}}{\p z} e^{i2t}\right\}\right]
\end{align}
again produces steady and oscillatory terms at order $\varepsilon^2$, with the steady portion relevant to the boundary condition \eqref{eq:general_bc}.
Expanding the boundary velocity term $\vu = \varepsilon \hU_k e^{it}$ yields at $\vx = \vx_k$
\begin{equation}
	\varepsilon \hvu e^{it} + \varepsilon^2 \left[\bvu - \frac{i\hU_k}{2} \frac{\p \hvu^*}{\p z}\right] = \varepsilon \hU_k e^{it} \zhat,
\end{equation}
giving the conditions
\begin{equation}
	\hvu = \hU_k \zhat, \qquad \bvu = \frac{i\hU_k}{2} \frac{\p \hvu^*}{\p z}
\end{equation}
on the static surface $\overline{\p\Omega}_k$.
The condition as $|\vx| \to \infty$ is translated in a more straightforward way as
\begin{equation}
	\varepsilon \hvu + \varepsilon^2 \bvu = -\varepsilon^2 \bU,
\end{equation}
to obtain the conditions
\begin{equation}
	\lim_{|\vx| \to \infty} \hvu = \vec{0}, \qquad \lim_{|\vx|\to\infty} \bvu = -\bU
\end{equation}
at the far point.

The fluid-mediated force on sphere $k$, to the orders accepted in our ans\"atz, is
\begin{equation}
	F_k^{(f)} = \int_{\overline{\p\Omega}_k} \nhat \bcdot \left[\varepsilon \htT e^{it} + \varepsilon^2 \left(\btT - \frac{i \hU_k}{2} \frac{\p \htT^*}{\p z}\right)\right] \bcdot \zhat dS. 
	\end{equation}
If we introduce the force fields
\begin{equation}
	\hF_k^{(f)} = \int_{\overline{\p\Omega}_k} \nhat \bcdot \htT \bcdot \zhat dS, \qquad \bF_k^{(f)} = \int_{\overline{\p\Omega}_k} \nhat \bcdot \btT \bcdot \zhat dS,
	\end{equation}
and the effective steady applied force on sphere $k$,
\begin{equation}
	\bF_k^{(a)} = - \frac{i \hU_k}{2} \int_{\overline{\p\Omega}_k} \nhat \bcdot \frac{\p \htT^*}{\p z} \bcdot \zhat dS,
	\end{equation}
then the fluid-mediated force takes the form
\begin{equation}
	F_k^{(f)} = \varepsilon \hF_k^{(f)} e^{it} + \varepsilon^2 \left(\bF_k^{(f)} + \bF_k^{(a)}\right).
	\end{equation}
However, $\bF^{(a)} = \bF_1^{(a)} + \bF_2^{(a)} = 0$, since by the divergence theorem (letting $\overline{\p\Omega} = \overline{\p\Omega}_1 \cup \overline{\p\Omega}_2$),
\begin{align}
\bF^{(a)} &= \int_{\overline{\p\Omega}} \nhat \bcdot \frac{\p \htT^*}{\p z} \bcdot \zhat dS \nonumber \\&= \int_\Omega \gr \bcdot \frac{\p\htT^*}{\p z} \bcdot \zhat dV \nonumber \\ &= \int_\Omega \frac{\p}{\p z}\left(\gr \bcdot \htT^* \bcdot \zhat\right) dV \nonumber \\
    &= \int_\Omega \frac{\p}{\p z}\left(\alpha^2 \hvu^* \bcdot \zhat\right)dV.
\end{align}
Here we have used $\gr \bcdot \tT = \alpha^2 \hvu$. After writing $\widehat{u}_z = \hvu \bcdot \zhat$, the equality $\bF^{(a)} = 0$ follows from writing
\begin{equation}
 \bF^{(a)} = \int_{-\infty}^\infty \int_{-\infty}^\infty D(x,y)  dx \ dy, \qquad D(x,y) = \int_{\shortstack{\scriptsize $z$ s.t. \\ \scriptsize $(x,y,z)\in \Omega$}} \alpha^2 \frac{\widehat{u}_z^*}{\p z} dz,
\end{equation}
where $D(x,y)$ corresponds to the integral of the exact derivative $\alpha^2 \p_z \widehat{u}_z^*$ over all $z$ in the fluid for the provided $x,y$. Since $\hvu_z \to \vec{0}$ as $z \to \pm \infty$ and is constant on both spheres, $D(x,y) = 0 \ \forall \ x,y$.

Finally, we note that $F_1^{(i)} = F_2^{(i)}$ and that by symmetry the time-averaged force applied to the fluid must vanish. This yields the conditions
\begin{equation}
	\hF_1^{(n)} - \hF_1^{(f)} = \hF_2^{(f)} - \hF_2^{(n)}, \qquad \bF_1^{(f)} + \bF_2^{(f)} = 0.
\end{equation}

\section{Calculation of swim speeds via the reciprocal theorem}\label{app:reciprocal}
In this section, we describe the Lorentz reciprocal theorem and show that it can be used to calculate the time-averaged swim speed of the dimer as a function of the leading-order oscillation described by the Brinkman amplitude field $(\hvu,\hp)$.

Consider two Stokes flows $(\bvu,\bp)$ and $(\vu',p')$ defined on the domain $\Omega$, driven by body forces $\bvf$ and $\vf'$, with associated traction vectors $\bvt = \overline{\tT} \bcdot \nhat$ and $\vt' = \tT'\bcdot \nhat$ on the time-averaged boundary $\overline{\p\Omega}$, where $\nhat$ points into the fluid (i.e.\@ the tractions correspond to forces applied to the spheres.) The generalised reciprocal theorem \citep{happel2012} requires the two flows satisfy
\begin{equation}
    \int_\Omega \bvu \bcdot \vf' \, dV - \int_{\overline{\p\Omega}} \bvu \bcdot \vt' \, dS = \int_\Omega \vu' \bcdot \bvf \, dV - \int_{\overline{\p\Omega}} \vu' \bcdot \bvt \, dS.
\label{eq:recip_base}
\end{equation}

Now, we let the barred flow represent our steady flow defined by \eqref{eq:Stokes}, shifted to the lab frame (instead of the swimmer frame.) We denote the surface velocity on sphere $k$ as $\bvu_s := (i \hU_k/2) \p_z \hvu^*$ (i.e.\@ $\bvu_s$ refers generally to the surface velocity on either sphere.) We define the primed flow as the one resulting from motion of the two-sphere system at a speed $U'$ under the influence of an applied force $F'$ in the absence of any body force.
With this assumption, $\vf' = \vec{0}$, and on the boundary of sphere $k$, $\vu' = U' \zhat$ and $\bvu = \overline{U} \zhat + \bvu_s$ as in  \eqref{eq:bcs}.
After making these substitutions, \eqref{eq:recip_base} simplifies to
\begin{equation}
    \bU F' =  \int_{\overline{\p\Omega}} \bvu_s \bcdot \vt' \, dS + \int_\Omega \vu' \bcdot \bvf \, dV,
\end{equation}
where we have used the fact that $\int_{\overline{\p\Omega}}\vu' \bcdot \bvt dS = 0$, and $\int_{\overline{\p\Omega}} \bvu \bcdot \vt' \,dS = -\bU F' + \int_{\p \Omega} \bvu_s \bcdot \vt' \,dS.$ Each of these terms can be understood in terms of physical effects.
The first integral corresponds to a slip velocity directed, on average, in one direction.
This yields motion in the other direction as in the case of squirmers \citep{Pedley16}, since to first order $\vt' \propto -\zhat$.
The second integral corresponds to a body force in the fluid that drives motion in the same direction in which it points.
Finally, note that
\begin{equation}
	\bvf = -(M^2/2) \bvu \bcdot \gr \bvu^* = -(M^2/2) \gr \bcdot(\bvu \otimes \bvu^*) = \gr \bcdot \btR
\end{equation}
as discussed in the main text.
We can integrate by parts to find
\begin{equation}
	\int_\Omega \vu' \bcdot \bvf dV = \int_{\overline{\p\Omega}} \nhat \bcdot \btR \bcdot \vu'\, dS - \int_\Omega (\gr \vu')\vec{:}\btR \,dV,
\end{equation}
but since $\btR \propto \zhat \otimes \zhat$ on sphere surfaces, the surface integral above vanishes.
Noting also that $\btR$ is symmetric, we have
\begin{equation}
    \bU F' =  \int_{\overline{\p\Omega}} \bvu_s \bcdot \vt' \, dS - \int_\Omega \te'\vec{:}\btR \, dV,
\end{equation}
which, on substituting back in for the definitions of $\btR$ and $\bvu_s$, gives \eqref{eq:swim_speeds}.

\section{Scalar, vector and tensor spherical harmonics}\label{app:harmonics}

\subsection{Spherical harmonics}

\subsubsection{Orthogonal bases for scalar, vector and tensor functions}
Let $(r,\theta,\phi)$ denote the usual spherical coordinates where $\theta$ is the polar angle measured from the positive $z$-axis.
We define the following inner products over the surface of the sphere for scalar fields (e.g.
$g=g(\vx), \ h=h(\vx)$), vectors $(\vg,\vh)$, and tensors $(\tg,\teh)$:
\begin{equation}
    \left<g,h\right> = \int_\Omega g h \ d\Omega, \qquad \left<\vg,\vh\right> = \int_\Omega \vg \bcdot \vh \ d\Omega, \qquad \left<\tg,\teh\right> = \int_\Omega \tg \boldsymbol{:} \teh \ d\Omega,
\end{equation}

We seek an orthogonal, axisymmetric set of basis functions for scalar, vector and tensor fields, which we will define in terms of spin-weighted spherical harmonics $Y_{ls}$, a generalisation of the Laplace harmonics which can be defined in terms of an axisymmetric spin-raising operator \citep{dray1985}.
Letting
\begin{equation}
    Y_{ls} = \mathcal{D}^s_+ Y_l, \qquad \mathcal{D}_\pm f := \left(\sin^{\pm s} \theta \right) \frac{\partial}{\partial \theta} \left[\left(\sin^{\mp s} \theta \right)f\right],
\end{equation}
then
\begin{equation}
	Y_{ls}(\theta) = (-1)^s \sqrt{\frac{2l+1}{4\pi}} \sin^s \theta \ P^{(s)}_l(\cos\theta),
\end{equation}
where $P^{(s)}_l(\cos \theta)$ is the $s$-th derivative of the Legendre polynomial $P_l$.
These satisfy
\begin{equation}
	\left<Y_{ls},Y_{jt}\right> = \frac{(l+s)!}{(l-s)!}\delta_{lj}\delta_{st}.
\end{equation}
Note the normalisation chosen here is different than the usual by a factor $\frac{(l+s)!}{(l-s)!}$, and that we include a factor $(-1)^s$.
This is so that $Y_{l1} = Y'_l(\theta)$.
Now, we can define a set of vector harmonics
\begin{equation}
	\vY_l^{(r)} = Y_{l0} \rhat, \qquad \vY_l^{(\theta)} = Y_{l1}(\theta) \that,
\end{equation}
which are orthogonal such that
\begin{equation}
\left<\vY_j^{(\beta)},\vY_k^{(\gamma)}\right> = \nu_{j,\beta} \delta_{jk} \delta_{\beta \gamma}, \label{eq:vec_inner_prod}
\end{equation}
with
\begin{equation}
	\nu_{l,r} = 1, \qquad \nu_{l,\theta} = l(l+1).
\end{equation}
Finally, we define a set of five rank-2 tensors using Cartesian unit vectors.
Let
\begin{equation}
	\tY_l^{(rr)} = Y_{l0} \ve_r \ve_r, \qquad \tY_l^{(\Omega_i)} = \frac12 Y_{l0} (\ve_\theta\ve_\theta+\ve_\phi\ve_\phi),
\end{equation}
describe axial and hoop isotropic tensors, both of which have nonzero traces, and let the traceless tensors
\begin{equation}
	\tY_l^{(r\theta)} = Y_{l1} \ve_r \ve_\theta, \qquad \tY_l^{(\theta r)} =  Y_{l1} \ve_\theta \ve_r, \qquad \tY_l^{(\Omega_s)} = \frac12 Y_{l2} (\ve_\theta\ve_\theta - \ve_\phi  \ve_\phi),
\end{equation}
describe radial--polar and hoop shear tensors.
This tensor basis is orthogonal, satisfying
\begin{equation}
	\left<\tY_j^{(\beta)},\tY_k^{(\gamma)}\right> = \kappa_{j,\beta} \delta_{jk}\delta_{\beta \gamma}, \label{eq:ten_inner_prod}
\end{equation}
with
\begin{equation}
	\kappa_{j,rr} = \kappa_{j,\Omega_i} = 1, \qquad \kappa_{j,r\theta} = \kappa_{j,\theta r} = l(l+1), \qquad \kappa_{j,\Omega_s} = \frac{(l+2)!}{(l-2)!}.
\end{equation}
These five are sufficient to describe the tensor fields we will encounter, since the assumption of axisymmetry precludes any azimuthal component to vector shear. 

We also consider a triple product between two vector fields $\vg,\vh$ and a tensor field $\tg$,
\begin{equation}
	\left<\vg, \tg, \vh\right> = \int_{\p\Omega} \vg \bcdot \tg \bcdot \vh \ dS.
\end{equation}
We define a quantity $N^{j,k,l}_{m,n,o}$ in therms of the Wigner 3$j$-symbols,
\begin{equation}
    N^{j,k,l}_{m,n,o} = \sqrt{\frac{(2j+1)(2k+1)(2l+1)}{4\pi}}\threeJ{j}{0}{k}{0}{l}{0}\threeJ{j}{m}{k}{n}{l}{o},
\end{equation}
which by symmetry is nonzero only if $j+k+l$ is even.
Then the basis harmonics satisfy
\begin{equation}
	\left<\vY_j^{(\beta)},\tY_k^{(\gamma)},\vY_l^{(\mu)}\right> = \Lambda_{j,k,l}^{(\beta,\gamma,\mu)},
\end{equation}
where the nonzero tensors elements are
\begin{equation}
	\Lambda_{j,k,l}^{(r,rr,r)} = N^{j,k,l}_{0,0,0},
\end{equation}
\begin{equation}
	\Lambda_{j,k,l}^{(\theta,\Omega_i,\theta)} = -\sqrt{\frac12\frac{(j+1)!}{(j-1)!}\frac{(l+1)!}{(l-1)!}}N^{j,k,l}_{1,0,-1},
\end{equation}
\begin{equation}
	\Lambda_{j,k,l}^{(r,r\theta,\theta)} = -\sqrt{\frac{(k+1)!}{(k-1)!}\frac{(l+1)!}{(l-1)!}} N^{j,k,l}_{0,1,-1},
\end{equation}
\begin{equation}
	\Lambda_{j,k,l}^{(\theta,\theta r,r)} = -\sqrt{\frac{(k+1)!}{(k-1)!}\frac{(j+1)!}{(j-1)!}}N^{j,k,l}_{-1,1,0},
\end{equation}
\begin{equation}
	\Lambda_{j,k,l}^{(\theta,\Omega_s,\theta)} = \sqrt{\frac12\frac{(j+1)!}{(j-1)!}\frac{(k+2)!}{(k-2)!}\frac{(l+1)!}{(l-1)!}} N^{j,k,l}_{-1,2,-1}.
\end{equation}

\subsection{Velocity field expansions}
Knowledge of these bases allows for easier calculation of the integrals in the swim speed equations \eqref{eq:swim_speeds}, but first, we consider expansions and derivative fields of a flow $(\vu,p)$ in terms of the scalar and vector harmonics above.
For some scalar functions $u_{r,l}(r)$ and $u_{\theta,l}(r)$, we have
\begin{equation}
	\vu(r,\theta) = \sum_{l=1}^\infty u_{r,l}(r) \vY_l^{(r)}(\theta) + u_{\theta,l}(r) \vY_l^{(\theta)}(\theta), \qquad p(r,\theta) = \sum_{l=1}^\infty p_l(r) Y_l(\theta).
\end{equation}
Let $H_{l;j} = (u_{r,l} - j u_{\theta,l})/r$.
Then the corresponding surface traction is
\begin{equation}
	\vt(r,\theta) = \sum_{l=1}^\infty \left[\left(2 u_{r,l}' - p_l\right)\vY_l^{(r)} + \left(H_{l;1} + u_{\theta,l}'\right) \vY_l^{(\theta)}\right],
\end{equation}
and the rate-of-strain tensor is
\begin{multline}
	\te(r,\theta) = \sum_{l=1}^\infty \Big[ u_{r,l}' \tY_l^{(rr)} + \sqrt{2} \ H_{l;l(l+1)/2} \tY_l^{(\Omega_i)} \\ + \frac12\left(H_{l;1} + u_{\theta,l}'\right)\left(\tY_l^{(r\theta)} + \tY_l^{(\theta r)}\right) + \frac{u_{\theta,l}}{\sqrt{2} r} \tY_l^{(\Omega_s)}\Big].
\end{multline}
Note that each mode of the surface traction and rate-of-strain tensor depend only on the same mode of the velocity field.
In contrast, the derivative in the $z$-direction is, letting $M_l = 1/\sqrt{(2l-1)(2l+1)}$,
\begin{multline}
	\frac{\p \vu}{\p z} = M_l \left[l  \left[(l+1)H_{l;1} + u_{r,l}'\right] \vY_{l-1}^{(r)} + \left[-H_{l;(l+1)^2} + (l+1)u_{\theta,l}'\right]\vY_{l-1}^{(\theta)}\right]\\
	+ M_{l+1} \left[(l+1)  \left[-l H_{l;1} + u_{r,l}'\right]\vY_{l+1}^{(r)} 
	+  \left[H_{l;l^2} + l u_{\theta,l}'\right]\vY_{l+1}^{(\theta)}\right], \label{eq:du_dz}
\end{multline}
so each mode $l$ of the vertical-derivative field depends on modes $l+1$ and $l-1$ of the velocity field.

\section{Addition theorems and reflection tensors}\label{app:reflect}

Recall from the main text that we consider reflected flows so that
\begin{equation}
\vu^{(k,r)} = \sum_{l=1}^{n_l} \sum_{m=0}^4 c^{(k)}_{lm} \vu_{lm},
\end{equation}
is the flow, written in harmonics about sphere $k$, at the $r$th step in the reflection.
We also decompose this into singular and regular parts
\begin{equation}
    \vu^{(-k,r)} = \sum_{l=1}^{n_l}\left(c_{l0}^{(k)} \vu_{l0} + c_{l1}^{(k)}\vu_{l1}\right), \qquad \vu^{(+k,r)} = \sum_{l=1}^{n_l} \left(c_{l2}^{(k)}\vu_{l2} + c_{l3}^{(k)}\vu_{l3}\right),
\end{equation}
and identify the corresponding set of coefficients as $c^{(-k)}_{lm} = \{c^{(k)}_{l0},c^{(k)}_{l1}\}$ and $c^{(+k)}_{lm} = \{c^{(k)}_{l2},c^{(k)}_{l3}\}$.
The total velocity field can thus be approximately expressed as
\begin{equation}
    \vu \approx \sum_{r = 0}^\infty \vu^{(1,r)} \approx \sum_{r=0}^\infty \vu^{(2,r)} \approx \sum_{r=0}^\infty \left(\vu^{(-1,r)}+\vu^{(-2,r)}\right).
\end{equation}
In other words, near sphere 1 or 2 we may consider the flow field in terms of a mixture of regular and singular modes centred around that sphere.
This is useful for evaluating integrals over sphere surfaces or in volumes closely surrounding them, but the presence of regular harmonics causes the expression to diverge as the distance from the origin approaches the sphere separation.

We may also consider the flow in terms of the sum of the singular fields originating around both spheres, which is more consistent with the physical description of each step in the reflection, wherein each rigid sphere induces a singular flow field to cancel out the influence of the other sphere's flow (represented at the location of the first sphere in terms of regular harmonics.) 
Thus, applying the method of reflections as in \figref{fig:reflections} requires two steps: first, one must describe a set of singular vector spherical harmonics originating at one sphere ($k$) in terms of regular harmonics about the other ($m$).
This is accomplished through the application of addition theorems which we detail in this section.
Second, one must describe the singular field which is induced at sphere $m$ by the presence of the flow originating at $k$.
We refer to the tensor describing these interactions as an ``reflection tensor.''

\subsection{Addition theorems}
Now, we seek to relate singular fields about one sphere to regular fields about the other.
This is achieved through the interaction tensor $A$, with
\begin{equation}
    c^{(+j)}_{lm} = A^{jk}_{lmno} c^{(-k)}_{no}.
\end{equation}
\renewcommand\vd{\vec{d}}
The elements can be derived from addition theorems for spherical harmonics.
Recall from \eqref{eq:Brinkman_basis} that $\vu_{l0} = \gr \phi_l^-$, where $\phi^-_l$ is the $l$th singular solid harmonic.
Since the flow depends linearly on the spherical harmonic, we can use addition theorems for the singular solid harmonics $f_l^-(\vx) = \phi^-_l(r) Y_l(\theta)$ which in the axisymmetric case are given by \begin{equation}
    \vu_{l0}(\vx + \vd) = \sum_{\lambda=0}^\infty  A^{(\phi)}_{l \lambda}(\vd) \vu_{\lambda2}(\vx).
\end{equation}
\begin{equation}
     A^{(\phi)}_{l\lambda}(\vd) = \frac{4\pi (-1)^\lambda}{2\lambda+1}
     {2l + 2\lambda \choose 2\lambda} {l+\lambda \choose \lambda}^{-1}
     f_{l+\lambda}^-(\vd) N^{l,\lambda,l+\lambda}_{0,0,0}.
\end{equation}
Here $\vx$ is a position vector in a spherical coordinate system about sphere $j$, and $\vd = \pm \bd \zhat$ is the displacement from sphere $k$ to the sphere $j$. 
Similarly, addition theorems exist for $\vu_{l1} = \gr \times \gr \times \left(\vx \chi_l^-\right)$ as wave solutions to the vector Helmholtz equation \citep{felderhof1987}.
Letting $f_l^-(\vx) = \chi_l^-(r) Y_l(\theta)$, this gives
\begin{equation}
    \vu_{l1}(\vx+\vd) = \sum_{\lambda=0}^\infty A^{(\chi)}_{l\lambda}(\vd) \vu_{\lambda3}(\vx),
\end{equation}
\begin{equation}
    A^{(\chi)}_{l\lambda}(\vd) = \sum_{\nu=0}^{\min\{l,\lambda\}} \frac{4\pi (-1)^\lambda}{\lambda(\lambda+1)} \left[2(l+\lambda-\nu)\nu-l\lambda\right] f_{l+\lambda-2\nu}^-(\vd{}) N^{l,\lambda,l+\lambda - 2\nu}_{0,0,0}.
\end{equation}
Using these definitions,
\begin{equation}
    A^{jk}_{l2m0} = A_{lm}^{(\phi)}(\vx_k - \vx_j), \qquad A^{jk}_{l3m1} = A_{lm}^{(\chi)}(\vx_k - \vx_j),
\end{equation}
where $\vx_m$ are the centres of sphere $m$.

\subsection{Induction and reflection tensors}
At the $n$th step in the reflection procedure, the presence of an incident flow at sphere $j$ originating at sphere $k$ is indicated by nonzero regular coefficients $c_{lm}^{(+j,n+1)}$.
In order to satisfy the velocity boundary condition on each sphere, we must find the singular coefficients $c_{lm}^{(-j,n+1)}$ corresponding to no flow on the boundary, so that $\vu_l^{(j,n+1)} = \vec{0}$ on $r = a_j$.
This can be accomplished by projecting the incident field $\vu_l^{(+j,n+1)}$ onto the singular basis $\{\vu_{l0},\vu_{l1}\}$.
The coefficients must therefore satisfy
\begin{equation}
    c_{lm}^{(-j,n)} = I_{lmno}^{j} c_{no}^{(+j,n)}, \qquad I_{lmno}^{j} = -\left<\vu_{lm},\vu_{no}\right>_j^{(a_j)},
\end{equation}
where $\left<*,*\right>_j^{(a)}$ describes the angular inner product \eqref{eq:vec_inner_prod} in the sphere $j$ coordinate system, with radial coordinate $r = a$.
Note this implies the only nonzero elements are $I^j_{lmln}$ for $m \in\{0,1\}$ and $n \in\{2,3\}$.
Recalling that $k_l = k_l(\alpha r)$ and $g_l = g_l(\alpha r)$ are the singular and regular modified spherical Bessel functions, these can be calculated as
\begin{equation}
    I_{l0l2}^{j} = \frac{l a_j^{2l+1}}{l+1} \frac{k_{l+1}}{k_{l-1}}, \qquad I_{l0l3}^{j} = \frac{2l+1}{l+1} \frac{r^{l-1}}{\alpha k_{l-1}}, 
\end{equation}
\begin{equation}
    I_{l0l3}^{j} = \frac{\alpha l r^{l+2}}{(2l+1)k_{l-1}} \left(g_{l-1}k_{l+1}-g_{l+1}k_{l-1}\right), \qquad I_{l1l3}^{j} = \frac{g_{l-1}}{k_{l-1}}.
\end{equation}
Combining these two relations gives us the total reflection tensor $R$ with
\begin{equation}
    c^{(-j,n)}_{lm} = R^{jk}_{lmpq} c^{(-k,n-1)}_{pq}, \qquad R^{jk}_{lmpq} = I^{j}_{lmno} A^{jk}_{nopq}.
\end{equation}

\section{Swim speed calculations}\label{app:speed_calcs}
In this section, we relate the integrals in \eqref{eq:swim_speeds} to inner products in the space of harmonics above.
In doing so, we associate the leading-order swim speed contributions with the appropriate interaction types in \figref{fig:amplitudes}.
The swim speeds are given by
 \begin{equation}
	\bU_b = \sum_{k=1}^2 \frac{i \hU_k}{2F'} \int_{\overline{\p \Omega_k}} \vt' \bcdot \frac{\p \hvu^*}{\p z} dS, \qquad \bU_r = \frac{M^2}{2F'} \int_\Omega \hvu \bcdot \te' \bcdot \hvu^* dV, \label{eq:swim_speeds_app}
\end{equation}
where $\vt' = \nhat \bcdot \tT'$ is the traction associated with the Stokes tow flow.
Let $\vu_l$, $\te_l'$ and $\vt_l'$ denote the $l$th harmonic contribution to the Brinkman velocity flow and Stokes rate-of-strain tensor and traction vector, respectively.
We will investigate the leading-order effects of the zeroth and first reflections.
At this order, the flow interaction corresponds to products of the $l=2$ Brinkman flow mode with the $l=1$ Stokes fields.
The geometric interaction corresponds to products of the $l=2$ modes of the virtual Stokes flow traction or strain with the $l=1$ Brinkman flow.

\subsection{Boundary velocity mechanism}
Note that $\bU_b$ in \eqref{eq:swim_speeds_app} consists of integrals over the sphere surfaces, and thus can be written exactly as
\begin{equation}
    \bU_b = \sum_{j,k=1}^{n_l}  \bU_b^{jk}, \qquad \bU_b^{jk} = \frac{i}{2 F'}\left( \hU_1 a_1^2 \left<\vt'_j, \frac{\p \hvu^*_k}{\p z}\right>_{1}^{(a_1)} +
    \hU_2 a_2^2 \left<\vt'_j, \frac{\p \hvu^*_k}{\p z}\right>_2^{(a_2)}\right),
\end{equation}
where $\left<*,*\right>_{k}^{(a)}$ denotes the inner product \eqref{eq:vec_inner_prod}, using an expansion in terms of spherical harmonics about sphere $k$, evaluated at the radial coordinate $r = a$ in the appropriate coordinate system.
With this convention, $\bU_b^{jk}$ represents the contribution from an interaction between the $j$th traction mode and $k$th velocity mode. 

\subsubsection{Low inertia}
At small $M^2$, the leading-order contributions to the swim speed are 
\begin{equation}
    \bU_b^{21} = \left[-\frac{3\sqrt{2} a^2_1 a^2_2(a_2-a_1)}{8(a_1+a_2)^3 \bd^2} + \bigO{1/\bd^3}\right]M + \bigO{M^3},
\end{equation}
corresponding to the geometric interaction, and
\begin{equation}
    \bU_b^{12} = 
        \left[\frac{3\sqrt{2}a_1^2 a_2^2(a_2-a_1)}{8(a_1+a_2)^3\bd^2} + \bigO{1/\bd^3}\right]M +
    \left[\frac{9a_1^2 a_2^2}{16(a_1+a_2)^3\bd} + \bigO{1/\bd^2}\right]M^2
    + \bigO{M^3},
\end{equation}
corresponding to the flow interaction ($\propto \hU_1 \hU_2^*$).
The dominant parts of the order-$M^2$ contribution from both cancel out, yielding the order-$M$ reported in the main text.

In the $M^2 = 0$ case using $S^2$ as a parameter, the leading order contribution is also from this interaction, and
\begin{equation}
    \bU_b^{12} = \left[-\frac{a_1 a_2(a_2 m_1 - a_1 m_2)}{3(a_1+a_2)^2 \bd^2} + \bigO{1/\bd^4}\right]S^2 + \bigO{S^4}.
\end{equation}

\subsubsection{High inertia}
At large $M^2$, the leading-order contributions to the swim speed are
\begin{equation}
    \bU^{21}_b = \left[\frac{3a_1 a_2(a_2 m_1^2 - a_1 m_2^2)}{4\sqrt{2}(a_1+a_2)(m_1+m_2)^2 \bd^2}\right]M + \bigO{1/M},
\end{equation}
from the geometric interaction, and
\begin{equation}
    \bU^{12}_b = \left[\frac{a_1^2 a_2^2(a_2 - a_1)m_1 m_2}{2\sqrt{2}(a_1+a_2)(m_1+m_2)^2 \bd^4}\right]M + \bigO{1/M},
\end{equation}
from the flow interaction.
As described in the main text, at high inertial levels the geometric interaction dominates the flow interaction, since outside of the width-$M^{-1}$ size boundary layer the Brinkman flow is weak.

In the $M^2 = 0$ Stokes case, as before the only contribution is from the flow interaction, which yields
\begin{equation}
    \bU^{12}_b = \left[\frac{27 a_1 a_2 (a_2 m_1 - a_1 m_2)}{4(a_1+a_2)(m_1+m_2)^2 \bd^2}\right]\frac{1}{S^2} + \bigO{\frac{1}{S^4}}.
\end{equation}

\subsection{Reynolds stress mechanism}

While the volume integral in \eqref{eq:swim_speeds_app} cannot be computed exactly in terms of angular integrals in the two spheres' coordinate systems, it can be computed approximately.
In the limit $a_1, a_2 \ll \bd$ and $M \ll 1$ or $M \gg 1$, we write
\begin{equation}
\bU_r \approx \bU_r^{jkl}, \qquad
	\bU_r^{jkl} \approx \frac{M^2}{2F'} \left(\int_{a_1}^\infty \left<\hvu_j,\te'_k,\hvu^*_l\right>_1^{(r)} r^2 dr + \int_{a_2}^\infty \left<\hvu_j,\te'_k,\hvu^*_l\right>_2^{(r)} r^2 dr \right), \label{eq:swim_speed_vh_inner_products}
\end{equation}
since the dominant contribution to the integrals in each case is concentrated near the sphere surfaces.
As before, we write the speed as a sum of contributions from interactions between particular harmonic modes, so that $\bU_r^{jkl}$ corresponds to the three-way interaction between the $j$-th and $l$-th flow modes and the $k$-th mode of the virtual Stokes rate-of-strain field.

The integrals over $r$ consist of sums of basis functions and flow coefficients, and some of these terms (corresponding to products of the regular basis functions) grow with $r$ and thus yield divergent integrals.
However, for the $l=1$ and 2 harmonic modes, it is possible by rescaling the two integrals' radial coordinates to combine them into a convergent integral.

\subsubsection{Low inertia}
For the flow interaction, we introduce $\bU_r^{(f)} = \bU_r^{112} + \bU_r^{211}$, which can be written in terms of two integrals
\begin{equation}
    \bU_r^{(f)} = \int_{a_1}^\infty f_1(r) dr + \int_{a_2}^\infty f_2(r) dr,
\end{equation}
for two functions $f_1$ and $f_2$.
For the first integral, we let $\xi = r/a_2$ and for the second $\xi = r/a_1$.
This gives rise to
\begin{equation}
    \bU_r^{(f)} = \int_{a_1/a_2}^\infty g_1(\xi) d\xi + \int_{a_2/a_1}^\infty g_2(\xi) d\xi,
\end{equation}
for $g_1(\xi) = a_2 f_1(\xi a_2)$ and $g_2(\xi) = a_1 f_2(\xi a_2)$.
In particular,
\begin{multline}
    g_1(\xi) = \\ \bigg[\frac{a_1^3(12 a_1^9 - 96 a_1^7 a_2^2 \xi^2 + 140 a_1^5 a_2^4 \xi^4 + 3a_1^4 a_2^5 \xi^5 -60 a_1^3 a_2^6 \xi^6 - 14 a_1^2 a_2^7 \xi^7 + 15 a_2^9 \xi^9}{80 a_2^6 (a_1+a_2)^3 \bd^2 \xi^9} \\ + \bigO{1/\bd^4}\bigg]M^2  + \bigO{M^3},
\end{multline}
and $g_2(\xi) = - g_1(\xi)|_{a_1 \leftrightarrow a_2}$, where the notation $\psi|_{x \leftrightarrow y}$ indicates that the symbols $x$ and $y$ are swapped within some expression $\psi$.
The only divergent term is the last in the numerator, which has the same magnitude and opposite sign in both functions.
Thus, we can write
\begin{equation}
    I_\infty = \int_{a_2/a_1}^\infty \left[g_1(\xi)+g_2(\xi)\right] d\xi, 
\end{equation}
and obtain a convergent integral, since the most slowly decaying term in the integrand is order $\xi^{-2}$.
After letting
\begin{equation}
    I_1 = \int_{a_1/a_2}^{a_2/a_1} g_1(\xi) d\xi
\end{equation}
to account for the portion of the first integral not included in $I_\infty$, we obtain
\begin{equation}
    \bU_r^{(f)} = I_\infty + I_1 = \left[\frac{15a_1^2 a_2^2(a_2-a_1)}{32(a_1+a_2)^2\bd^2} + \bigO{1/\bd^4}\right]M^2 + \bigO{M^3}.
\end{equation}

The geometric interaction is represented by a single contribution
\begin{equation}
    \bU_r^{121} = \int_{a_1}^\infty f_1(r) dr + \int_{a_2}^\infty f_2(r) dr,
\end{equation}
for two functions $f_1$ and $f_2$.
Now, we let $\xi = r/a_1$ for the first integral and $\xi = r/a_2$ for the second.
This yields
\begin{equation}
    \bU_r^{121} = \int_{1}^\infty \left[g_1(\xi) + g_2(\xi)\right] d\xi, \label{eq:geom_reyn_speed}
\end{equation}
where $g_k(\xi) = a_k f_k(\xi a_k)$.
Here,
\begin{multline}
    g_1(\xi) =  \bigg[\frac{a_1^3 a_2^3 \left(-36+97\xi^2-106\xi^4+\xi^5 +45\xi^6 -2\xi^7 +81 \xi^9\right)}{160(a_1+a_2)^3 \bd^2 \xi^9} \\ +\bigO{1/\bd^4}\bigg]M^2 + \bigO{M^3},
\end{multline}
and $g_2(\xi) = -g_1(\xi)$.
Thus, at this order $\bU_r^{121} \approx 0$ and there is no contribution from the geometric interaction, since the two integrals diverge at corresponding oppositely signed rates.

\subsubsection{High inertia}
In the high-inertia limit, the Stokeslet-like portion of the Brinkman flow is confined to a thin boundary layer around each sphere.
Thus, the flow interactions are weak, and $\bU_r^{(f)}$ is $\bigO{1/\bd^4}$.

For the geometric interaction, we again obtain a solution of the form of \eqref{eq:geom_reyn_speed}.
We can make a further substitution $\zeta = M(\xi-1)$, so that the small coordinate $\zeta$ corresponds to progress through the boundary layer.
We find that at leading order,
\begin{equation}
    g_1(\xi) + g_2(\xi) = g_\text{in}(\zeta) + g_\text{out}(\xi), \qquad \lim_{\zeta \to \infty} g_\text{in}(\zeta) = 0.
\end{equation}
Thus, we write the integral \eqref{eq:geom_reyn_speed} in terms of two integrals,
\begin{equation}
    \bU_r^{121} \approx \int_0^\infty g_\text{in}(\zeta) d\zeta + \int_1^\infty g_\text{out}(\xi) d\xi,
\end{equation}
where the integrands can be calculated as
\begin{multline}
	g_\text{in}(\zeta) = \\ \Bigg[\frac{3a_1 a_2\left(a_1^2 m_2^2 e^{-a_1 \zeta/\sqrt{2}} \cos\left(\frac{a_1 \zeta}{\sqrt{2}}\right) - a_2^2 m_1^2 e^{-a_2\zeta/\sqrt{2}} \cos\left(\frac{a_2 \zeta}{\sqrt{2}}\right)\right)}{4(a_1+a_2)(m_1+m_2)^2 \bd^2} + \bigO{1/\bd^4}\Bigg]M \\ + \Bigg[\frac{3 a_1 a_2}{4\sqrt{2}(a_1+a_2)(m_a+m_2)^2}\Bigg(\frac{a_2 m_1^2}{d^2} e^{-\frac{a_2 \zeta}{\sqrt{2}}}\left[3 e^{-\frac{a_2 \zeta}{\sqrt{2}}} - 5  \cos\left(\frac{a_2 \zeta}{\sqrt{2}}\right) - \sin\left(\frac{a_2 \zeta}{\sqrt{2}}\right)\right]  \\  - \frac{a_1 m_2^2}{d^2} e^{-\frac{a_1\zeta}{\sqrt{2}}} \left[3 e^{-\frac{a_1 \zeta}{\sqrt{2}}} - 5  \cos\left(\frac{a_1 \zeta}{\sqrt{2}}\right) - \sin\left(\frac{a_1 \zeta}{\sqrt{2}}\right)\right]\Bigg) + \bigO{1/d^4}\Bigg] + \bigO{1/M},
\end{multline}
\begin{equation}
    g_\text{out}(\xi) = \left[\frac{a_1 a_2(a_1^2 m_2^2 - a_2^2 m_1^2)(-36+25\xi^2 + \xi^5)}{40(a_1+a_2)(m_1+m_2)^2 \xi^9 \bd^2} + \bigO{1/\bd^4}\right]M^2 + \bigO{M}.
\end{equation}
At this order the outer integral $\int_{1}^\infty g_\text{out}(\xi)d\xi = 0$.
Thus, the contribution comes from within the boundary layer, and
\begin{multline}
    \bU_r^{121} = \int_0^\infty g(\zeta) d\zeta = \left[\frac{3 a_1 a_2 (a_1 m_2^2 - a_2 m_1^2)}{4\sqrt{2}(a_1 + a_2) (m_1+m_2)^2 \bd^2} + \bigO{1/\bd^4}\right]M  \\
	+\left[\frac{9 a_1 a_2 (m_2 - m_1)}{8(a_1 + a_2)(m_1+m_2) d^2} + \bigO{1/d^4}\right] + 
	\bigO{1/M}.
\end{multline}

\bibliographystyle{jfm}
\bibliography{streaming}

\end{document}